\documentclass[12pt,a4paper,dvipsnames]{amsart}
\usepackage[utf8]{inputenc}
\usepackage{tabularx}
\usepackage[table]{xcolor}
\usepackage{amsmath, nicefrac, amsthm, verbatim, amsfonts, amssymb, xcolor, bbm}
\usepackage{graphics, xspace, enumerate, cite}
\usepackage[a4paper,margin=3cm]{geometry}
\usepackage{mathrsfs}

\usepackage[bb=boondox]{mathalfa}
\usepackage{tikz}
\usepackage{graphicx}
\usepackage[colorlinks=true,citecolor=green,urlcolor=green,linkcolor=green,bookmarksopen=true,unicode=true,pdffitwindow=true]{hyperref}
\usepackage[english]{babel}
\usepackage[languagenames,fixlanguage]{babelbib}

\usepackage[labelformat=simple]{subcaption}

\usetikzlibrary{decorations.pathreplacing,calligraphy}

\makeatletter
\@namedef{subjclassname@2020}{\textup{2020} Mathematics Subject Classification}
\makeatother

\theoremstyle{plain}
\newtheorem{theorem}{Theorem}[section]

\theoremstyle{definition}
\newtheorem{remark}[theorem]{Remark}

\usepackage{textcomp}

\newcommand{\bbN}{\mathbb{N}}

\numberwithin{equation}{section}

\definecolor{Maroon}{RGB}{140,10,0}

\hypersetup{
	colorlinks,
	linkcolor={Maroon},
	citecolor={Maroon},
	urlcolor={Maroon}
}

\newcommand{\e}{\ensuremath{\text{\textopenbullet}}}
\newcommand{\f}{\ensuremath{\text{\textbullet}}}
\newcommand{\x}{\ensuremath{\text{\texttimes}}}

\graphicspath{{./figs/}}

\title{A model of random sequential adsorption on a ladder graph}

\author[T.\ Do\v{s}li\'{c}]{Tomislav\ Do\v{s}li\'{c}}
\address[Tomislav\ Do\v{s}li\'{c}]{University of Zagreb Faculty of Civil Engineering\\
	Department of Mathematics\\
	Fra Andrije Kačića-Miošića 26, 10000 Zagreb, Croatia \\ and
	Faculty of Information Studies \\
	Novo Mesto \\
	Slovenia}
\email{tomislav.doslic@grad.unizg.hr}

\author[M.\ Puljiz]{Mate\ Puljiz$^{\ast}$}
\thanks{$^\ast$Author to whom any correspondence should be addressed.}
\address[Mate Puljiz]{University of Zagreb Faculty of Electrical Engineering and Computing\\
	Department of Applied Mathematics\\
	Unska 3, 10000 Zagreb, Croatia}
\email{mate.puljiz@fer.unizg.hr}

\author[S.\ \v{S}ebek]{Stjepan\ \v{S}ebek}
\address[Stjepan\ \v{S}ebek]{University of Zagreb Faculty of Electrical Engineering and Computing\\
	Department of Applied Mathematics\\
	Unska 3, 10000 Zagreb, Croatia}
\email{stjepan.sebek@fer.unizg.hr}

\author[J.\ \v{Z}ubrini\'{c}]{Josip\ \v{Z}ubrini\'{c}}
\address[Josip\ \v{Z}ubrini\'{c}]{University of Zagreb Faculty of Electrical Engineering and Computing\\
	Department of Applied Mathematics\\
	Unska 3, 10000 Zagreb, Croatia}
\email{josip.zubrinic@fer.unizg.hr}


\begin{document}

\begin{abstract}
In random sequential adsorption (RSA), objects are deposited on a substrate randomly, irreversibly, and sequentially. Attempts of deposition that lead to an overlap with previously deposited objects are discarded. The process continues until the system reaches a jammed state when no further additions are possible.  We analyze a class of RSA models on a two-row square ladder graph in which landing on an empty site in a graph is allowed when at least $b$ neighboring sites in the graph are unoccupied ($b \in \bbN$). In this paper we complement this typical way of studying RSA models by analyzing also the structure of the set of all jammed states in a static way, disregarding the dynamics that led to a particular jammed state. In both considered settings (dynamic and static) we provide explicit expressions for key statistics that describe the average proportion of the substrate covered by deposited objects, and then we comment on significant differences between the two settings. We illustrate all of our findings through a toy model for ensembles of trapped Rydberg atoms with blockade range $b$.
\end{abstract}

\maketitle


%
%
%
%
\section{Introduction}

Random sequential adsorption (RSA) is a toy model mimicking the irreversible deposition of suspended particles onto substrates. Particles are deposited randomly, and sequentially, obeying the rule that if the new particle is sufficiently far away from already deposited ones, it sticks to the substrate; otherwise, the deposition event is discarded. In the two-dimensional setting, the RSA models have been applied to modeling chemisorption on single-crystal surfaces and adsorption in colloidal systems \cite{Evans1993, Talbot2000, Torquato2002, Adamczyk2017, Bressloff2013}. There are also applications in nanotechnology; see Refs.\ \cite{Chen2002, Elimelech2003, Gray2006, Katsman2013}. The RSA models have been also used in high dimensions, e.g., in the context of packing problems \cite{Torquato2010}.

The first and most famous RSA model was introduced by Flory \cite{Flory1939}. The purpose of his model was the description of reactions along a long polymer chain (using analogy with adsorption of dimers). Another beautiful RSA model with adsorption on a continuous one-dimensional line was introduced by Rényi \cite{Renyi1958} as a toy model of car parking. The class of RSA models has also been used in several other one-dimensional settings, e.g., in modeling polymer translocation \cite{D'Orsogna2007, Krapivsky2010}, and describing zero-temperature dynamics of Ising chains \cite{Privman1992, Krapivsky1994,DeSmedt2002}.

We illustrate the particular RSA model studied in this paper through ensembles of trapped Rydberg atoms. Our model mimics the generation of Rydberg excitations. Neutral atoms excited into a high-energy state, the so-called Rydberg state, have been intensely studied, and have become a testing ground for various quantum mechanical problems in quantum information processing, quantum computation, and quantum simulation \cite{Saffman2010}. In experiments (see \cite{Jaksch2000, Liebisch2005, Viteau2012, Malossi2014}), a laser excites ultra-cold atoms into a Rydberg state. Interactions between Rydberg atoms cause the blockage forbidding the excitation of atoms sufficiently close to a Rydberg atom. When the radiative decay of the Rydberg atoms can be ignored, this RSA model mimics certain features of the excitation process \cite{Sanders2014}, although it disregards features like the nonergodic quantum dynamics of Rydberg-blockaded chains (see \cite{Bernien2017, Turner2018, Ho2019, Khemani2019} and references therein). It is essential to note that this blockade is not absolute; under specific conditions, such as when the laser field is detuned, the anti-blockade phenomenon can occur. In this regime, atoms within the blockade radius can still be excited, allowing for greater control over excitation dynamics. Our toy model completely disregards this anti-blockade phenomenon.

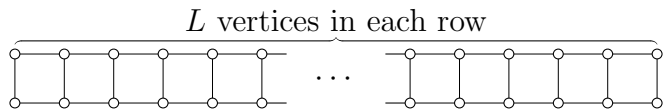
\begin{figure}
	\centering
	\begin{tikzpicture}[scale = 0.65]
		\draw[step=1cm,black,very thin] (0, 0) grid (5,1);
		\draw[step=1cm,black,very thin] (8, 0) grid (13,1);

		\draw (5, 0) -- (5.5, 0);
		\draw (5, 1) -- (5.5, 1);
		\draw (8, 0) -- (7.5, 0);
		\draw (8, 1) -- (7.5, 1);

		\node at (6.5, 0.5) {$\cdots$};

		\foreach \x in {0,...,5}{
			\draw[color=black, fill=white] (\x,0) circle [radius=0.1];
			\draw[color=black, fill=white] (\x,1) circle [radius=0.1];}

		\foreach \x in {8,...,13}{
			\draw[color=black, fill=white] (\x,0) circle [radius=0.1];
			\draw[color=black, fill=white] (\x,1) circle [radius=0.1];}

      \draw [decorate, decoration = {calligraphic brace}] (0,1.2) -- (13,1.2);
      \node at (6.5, 1.7) {$L$ vertices in each row};
	\end{tikzpicture}
	\caption{Two-row ladder graph.}
	\label{fig:ladder}
\end{figure}

\begin{figure}
	\centering
	\begin{subfigure}{0.4\textwidth}
		\centering
		\begin{tikzpicture}[scale = 0.65]
			\draw[step=1cm,black,very thin] (0, 0) grid (6,1);
			\foreach \x in {0,...,6}{
				\draw[color=black, fill=white] (\x,0) circle [radius=0.1];
				\draw[color=black, fill=white] (\x,1) circle [radius=0.1];}
		\end{tikzpicture}
		\caption{Non-jammed configuration with no excited atoms.}
		\label{fig:nJS0}
	\end{subfigure}\hfill
	\begin{subfigure}{0.4\textwidth}
		\centering
		\begin{tikzpicture}[scale = 0.65]
			\draw[step=1cm,black,very thin] (0, 0) grid (6,1);
			\foreach \x in {0,...,6}{
				\draw[color=black, fill=white] (\x,0) circle [radius=0.1];
				\draw[color=black, fill=white] (\x,1) circle [radius=0.1];}
			\draw[color=black, fill=black] (6,0) circle [radius=0.1];
		\end{tikzpicture}
		\caption{Non-jammed configuration with $1$ excited atom.}
		\label{fig:nJS1}
	\end{subfigure}
	\begin{subfigure}{0.4\textwidth}
		\centering
		\begin{tikzpicture}[scale = 0.65]
			\draw[step=1cm,black,very thin] (0, 0) grid (6,1);
			\foreach \x in {0,...,6}{
				\draw[color=black, fill=white] (\x,0) circle [radius=0.1];
				\draw[color=black, fill=white] (\x,1) circle [radius=0.1];}
			\draw[color=black, fill=black] (1,1) circle [radius=0.1];
			\draw[color=black, fill=black] (6,0) circle [radius=0.1];
		\end{tikzpicture}
		\caption{Non-jammed configuration with $2$ excited atoms.}
		\label{fig:nJS2}
	\end{subfigure}\hfill
	\begin{subfigure}{0.4\textwidth}
		\centering
		\begin{tikzpicture}[scale = 0.65]
			\draw[step=1cm,black,very thin] (0, 0) grid (6,1);
			\foreach \x in {0,...,6}{
				\draw[color=black, fill=white] (\x,0) circle [radius=0.1];
				\draw[color=black, fill=white] (\x,1) circle [radius=0.1];}
			\draw[color=black, fill=black] (1,1) circle [radius=0.1];
			\draw[color=black, fill=black] (4,1) circle [radius=0.1];
			\draw[color=black, fill=black] (6,0) circle [radius=0.1];
		\end{tikzpicture}
		\caption{Jammed configuration with $3$ excited atoms.}
		\label{fig:JS}
	\end{subfigure}
	\caption{One possible excitation sequence resulting in a jammed configuration of Rydberg atoms on the ladder of length $L = 7$, when blockade range $b = 2$.}
	\label{fig:jammed_state}
\end{figure}
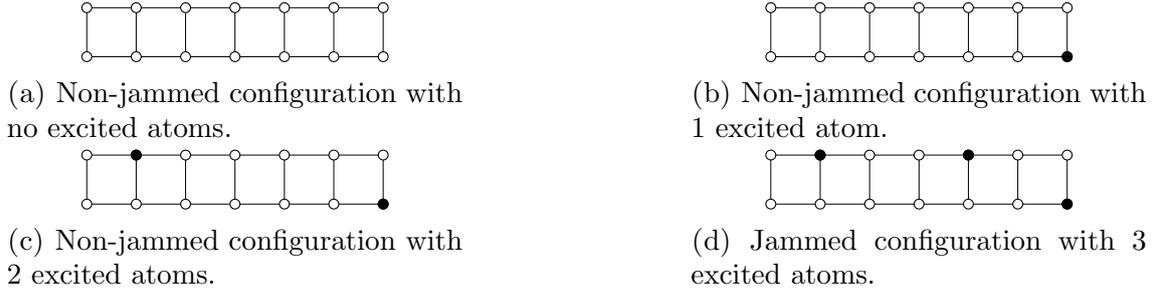

In our model, atoms occupy vertices of a two-row square ladder graph of length $L$ with $2L$ vertices (see Figure \ref{fig:ladder}). We start with all the atoms in a neutral state (represented by unfilled vertices), and then we randomly, and sequentially excite them into a Rydberg state (represented by filled vertices). The blockage effect is modeled by the requirement that excited atoms are separated by at least $b$ neutral atoms (where $b\in \bbN$ is a parameter of our model). More precisely, the shortest-path distance between any two excited atoms must be at least $b+1$. We proceed with random sequential excitation of neutral atoms into a Rydberg state until we find ourselves in a situation where it is impossible to excite another neutral atom without violating the blockage constraint (i.e.\ exciting any of the neutral atoms would result in two excited atoms being strictly less than $b+1$ edges apart). Figure \ref{fig:jammed_state} shows one possible sequence of random sequential excitations on the ladder of length $L=7$ when blockade range is $b=2$.

A configuration of atoms, on our underlying ladder graph, for which it holds that it is impossible to excite another neutral atom without violating the blockage constraint is called \emph{jammed (or maximal) configuration}. For example, the configuration in Figure \ref{fig:JS} is jammed because exciting any other atom, in addition to the three already excited, would result in that fourth excited atom being at a (edge) distance less than $3=2+1$ from another excited atom. Note that, depending on a particular order in which atoms are excited, a jammed configuration reached at the end of the process can have different number of atoms in a Rydberg state (see Figure \ref{fig:JC_on_L=7} below).

It is worth mentioning again that this model completely disregards the anti-blockade phenomenon which would allow atoms within the blockade radius to be excited, thus allowing for greater control over excitation dynamics. Disregarding the anti-blockade phenomenon is crucial for our model, as the notion of jammed configurations only makes sense if the excitations are irreversible and the blockage constraint is strictly enforced.

We are interested in the proportion (density) of the excited atoms in jammed configurations, which is defined as the ratio of the number of excited atoms, and the total number of atoms we started with (i.e.\ the total number of vertices in our ladder graph). The main question related to jammed configurations concerns finding the expected proportion (density) of the excited atoms. More precisely, if we sample a single jammed configuration from the set of all the jammed configurations, what is the expected value of the proportion of the excited atoms in the sampled jammed configuration. Clearly, the answer to that question depends on the way we sample a single jammed configuration from the set of all jammed configurations.

There are two natural ways of doing that. One way is through the RSA process (explained above). We randomly and sequentially excite atoms into a Rydberg state until we reach a jammed configuration. We refer to this way of sampling one jammed configuration as the \emph{dynamic model}. The expected density of the excited atoms in this setting is called the \emph{jamming limit}. To determine the value of the jamming limit one can run experiments (see \cite{Marvel1938}), computer simulations (see \cite{Nord1991, Tory1983, Wang2000, Georgiou2009}), or use analytical methods (see \cite{Flory1939, Renyi1958, Krapivsky2010a, Fan1992, Baram1992, Chern2015, Georgiou2009}). Even though several analytic solutions have been found, for most of the models studied in the literature, the jamming limits have only been approximated using computer simulations. Furthermore, most of the papers in which analytic solutions have been obtained are dealing with one-dimensional models. In this paper we study a very modest generalization of the one-dimensional lattice, namely the two row square ladder graph (see Figure \ref{fig:ladder}), but we manage to obtain the analytic solution for all $b \ge 1$ (where the parameter $b$ models the blockade range of atoms in a Rydberg state).

Another way to look at the problem of the expected density of the excited atoms in a jammed configuration is to assume that all the jammed configurations in our model are equally likely to appear, and then sample one such configuration at random. This approach is usually referred to as the \emph{equilibrium (or static) model}. Different equilibrium models have also been studied in the literature (see \cite{Crisanti2000, DeSmedt2002, Dean2000, Krapivsky2010a, Lefevre2001, Lefevre2001a}). The standard way to describe the equilibrium model is by the so-called complexity function (also known as configurational entropy). It is known that in similar models, the number of different jammed configurations with a particular density tends to grow exponentially with the length of the configuration (see \cite{Palmer1985, Elskens1987, Krapivsky1994}). The complexity function then describes (in a refined way) what portion of the total number of configurations is taken up by configurations having a particular density. More precisely, the complexity function assigns to each density $\rho$ the exponential rate $S(\rho)$ at which the number of configurations with density $\rho$ grows as the length of the ladder increases to infinity. It then follows that the number of configurations with density $\rho$ is approximately $e^{LS(\rho)}$ for large ladder lengths $L$. As a consequence, the expected density of excited particles in a jammed configuration (in the equilibrium version of the model) approaches the argument of the maximum of the complexity function, as the length of configurations grows. This density at which the complexity function attains its maximum is called the \emph{equilibrium density}. More details on the intuition behind the definition of the complexity function, and on the possible ways of obtaining the explicit expressions for a complexity function are given in \cite{Puljiz2023b, Krapivsky2023}.

Initially, it is not clear that the two models (dynamic and equilibrium) lead to different distributions of jammed configurations. The assumption that both approaches result in the same distribution of jammed configurations is referred to as the Edwards hypothesis, see \cite{Baule2018} for a recent review. In the models studied in this paper (as in most cases), the Edwards hypothesis is violated.

The simplest setting for studying Rydberg atoms and their blockage effect is on a finite one-dimensional lattice. This model has already been considered in the literature. The jamming limit in the dynamic version of that problem was obtained in \cite{Friedman1964, Krapivsky2020, Mackenzie1962}, and the complexity function of the equilibrium model was recently obtained in \cite{Krapivsky2023, Doslic2024a}. In this paper, we use similar techniques to extend these results from the one-dimensional lattice, to a model of Rydberg atoms on the two-row square ladder graph shown in Figure \ref{fig:ladder}. In the rest of the paper we refer to this graph as the ladder.

The dynamic model of Rydberg atoms on the ladder has already been studied in the literature under different guises, but only for blockade range $b = 1$. In physics literature, it was studied as a generalization of the famous Flory model introduced in \cite{Flory1939}, where the imposed constraint is precisely the nearest neighbor exclusion (see \cite{Fan1992, Baram1992}), and in the mathematics literature, it was studied as the unfriendly theater seating arrangement problem \cite{Georgiou2009}, and random unfriendly seating arrangement at a dining table \cite{Friedman1964,Chern2015}. The equilibrium model for $b = 1$ is treated in \cite{Puljiz2023b}.

It is worth noticing that somewhat frivolous and caricatural settings of
unfriendly seating arrangements, or even more frivolous ``urinal problem''
\cite{Kranakis2010}, suddenly gained on relevance and respectability with the
outbreak of COVID pandemics in early 2020.

The rest of the paper is organized as follows. In Section \ref{sec:dynamic_model} we study the dynamic model and we obtain a closed formula for the jamming limit of the model of Rydberg atoms on the ladder, for arbitrary blockade range $b \ge 1$. Section \ref{sec:equilibrium_model} contains the analysis of the equilibrium model. The main result in that section is the expression for the complexity function of the model of Rydberg atoms on the ladder for all $b \ge 1$. In Section \ref{sec:comparison} we compare the two models. We show that the Edwards hypothesis is violated, and that the behavior of the two models differs quite significantly as $b$ tends to infinity. Finally, in Section \ref{sec:conclusion} we recapitulate our findings and indicate several possible directions of future research.

\section{Dynamic model}\label{sec:dynamic_model}
\begin{figure}
\begin{subfigure}{1.0\textwidth}
    \centering
    \begin{tikzpicture}[scale = 0.65]
            \draw[step=1cm,black,very thin] (0, 0) grid (3,1);
		\draw (0, 1) -- (-1, 1);
		\draw (3, 0) -- (4, 0);
            \draw (3, 1) -- (3.4, 1);
		\draw (4, 0) -- (4.4, 0);
            \draw (4, 0) -- (4, 0.4);
            \draw (5, 1) -- (4.6, 1);
            \draw (5, 1) -- (5, 0.6);
		\draw (6, 0) -- (5.6, 0);
		\foreach \x in {0,...,3}{
			\draw[color=black, fill=white] (\x,0) circle [radius=0.1];
			\draw[color=black, fill=white] (\x,1) circle [radius=0.1];}
            \draw[color=black, fill=white] (-1,1) circle [radius=0.1];
            \draw[color=black, fill=white] (4,0) circle [radius=0.1];
            \node[rectangle, draw=none, minimum size=1pt] at (4, 1) {$\cdots$};
            \node[rectangle, draw=none, minimum size=1pt] at (5, 0) {$\cdots$};
            \draw[step=1cm,black,very thin] (6, 0) grid (9,1);
		\draw (6, 1) -- (5, 1);
		\draw (9, 0) -- (10, 0);
		\foreach \x in {6,...,9}{
			\draw[color=black, fill=white] (\x,0) circle [radius=0.1];
			\draw[color=black, fill=white] (\x,1) circle [radius=0.1];}
            \draw[color=black, fill=white] (5,1) circle [radius=0.1];
            \draw[color=black, fill=white] (10,0) circle [radius=0.1];
            \draw [decorate, decoration = {calligraphic brace}] (-1.2,1.2) -- (9.2,1.2);
            \node[rectangle, draw=none, minimum size=1pt] at (4, 1.8) {$k$};
    \end{tikzpicture}
    \caption{$\mathscr{A}_k$ graph.}
    \label{fig:Ak}
\end{subfigure}

\bigskip

\begin{subfigure}{1.0\textwidth}
    \centering
    \begin{tikzpicture}[scale = 0.65]
            \draw[step=1cm,black,very thin] (0, 0) grid (4,1);
		\draw (0, 0) -- (-1, 0);
            \draw (4, 0) -- (4.4, 0);
            \draw (4, 1) -- (4.4, 1);
		\draw (6, 0) -- (5.6, 0);
            \draw (6, 1) -- (5.6, 1);
		\draw (9, 0) -- (10, 0);
		\foreach \x in {0,...,4}{
			\draw[color=black, fill=white] (\x,0) circle [radius=0.1];
			\draw[color=black, fill=white] (\x,1) circle [radius=0.1];}
            \draw[color=black, fill=white] (-1,0) circle [radius=0.1];
            \node[rectangle, draw=none, minimum size=1pt] at (5, 0) {$\cdots$};
            \node[rectangle, draw=none, minimum size=1pt] at (5, 1) {$\cdots$};
            \draw[step=1cm,black,very thin] (6, 0) grid (9,1);
		\foreach \x in {6,...,9}{
			\draw[color=black, fill=white] (\x,0) circle [radius=0.1];
			\draw[color=black, fill=white] (\x,1) circle [radius=0.1];}
            \draw[color=black, fill=white] (10,0) circle [radius=0.1];
            \draw [decorate, decoration = {calligraphic brace}] (-0.2,1.2) -- (9.2,1.2);
            \node[rectangle, draw=none, minimum size=1pt] at (5, 1.8) {$k-1$};
    \end{tikzpicture}
    \caption{$\mathscr{B}_k$ graph.}
    \label{fig:Bk}
\end{subfigure}
\caption{Subgraphs of the original ladder graph that appear, during the process of random sequential excitation of neutral atoms to a Rydberg state, due to the blockage effect of excited atoms.}
\label{fig:Ak_Bk}
\end{figure}
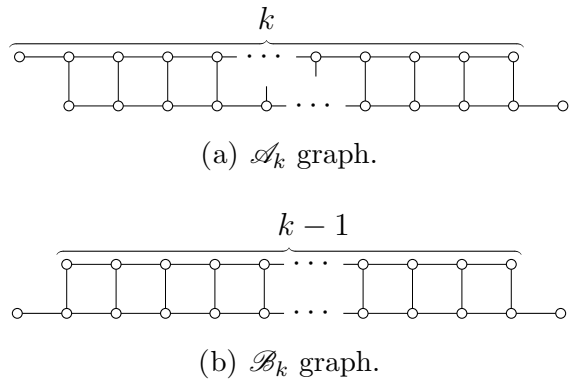

\begin{figure}
\centering
\begin{subfigure}{0.3\textwidth}
    \centering
    \begin{tikzpicture}[scale = 0.65]
		\draw (0, 0) -- (0, 1);
		\draw (0, 1) -- (-1, 1);
            \draw (0, 0) -- (1, 0);
            \draw[color=black, fill=white] (-1,1) circle [radius=0.1];
            \draw[color=black, fill=white] (0,1) circle [radius=0.1];
            \draw[color=black, fill=white] (0,0) circle [radius=0.1];
            \draw[color=black, fill=white] (1,0) circle [radius=0.1];
    \end{tikzpicture}
    \caption{$\mathscr{A}_2$}
    \label{fig:A2}
\end{subfigure}
\begin{subfigure}{0.3\textwidth}
    \centering
    \begin{tikzpicture}[scale = 0.65]
        \draw[color=black, fill=white] (-1,1) circle [radius=0.1];
        \draw[color=black, fill=white] (0,0) circle [radius=0.1];
    \end{tikzpicture}
    \caption{$\mathscr{A}_1$}
    \label{fig:A1}
\end{subfigure}
\begin{subfigure}{0.3\textwidth}
    \centering
    \begin{tikzpicture}[scale = 0.65]
        \node[rectangle, draw=none, minimum size=1pt] at (0, 1) {$\emptyset$};
    \end{tikzpicture}
    \caption{$\mathscr{A}_0$}
    \label{fig:A0}
\end{subfigure}

\bigskip

\begin{subfigure}{0.3\textwidth}
    \centering
    \begin{tikzpicture}[scale = 0.65]
		\draw (0, 0) -- (-1, 0);
		\draw (0, 0) -- (1, 0);
            \draw (0, 0) -- (0, 1);
            \draw[color=black, fill=white] (-1,0) circle [radius=0.1];
            \draw[color=black, fill=white] (0,0) circle [radius=0.1];
            \draw[color=black, fill=white] (1,0) circle [radius=0.1];
            \draw[color=black, fill=white] (0,1) circle [radius=0.1];
    \end{tikzpicture}
    \caption{$\mathscr{B}_2$}
    \label{fig:B2}
\end{subfigure}
\begin{subfigure}{0.3\textwidth}
    \centering
    \begin{tikzpicture}[scale = 0.65]
        \draw (0, 0) -- (1, 0);
        \draw[color=black, fill=white] (0,0) circle [radius=0.1];
        \draw[color=black, fill=white] (1,0) circle [radius=0.1];
    \end{tikzpicture}
    \caption{$\mathscr{B}_1$}
    \label{fig:B1}
\end{subfigure}
\begin{subfigure}{0.3\textwidth}
    \centering
    \begin{tikzpicture}[scale = 0.65]
        \draw[color=black, fill=white] (0,0) circle [radius=0.1];
    \end{tikzpicture}
    \caption{$\mathscr{B}_0$}
    \label{fig:b0}
\end{subfigure}
\caption{The graphs $\mathscr{A}_k$ and $\mathscr{B}_k$ for $k \in \{0, 1, 2\}$.}
\label{fig:k=012}
\end{figure}
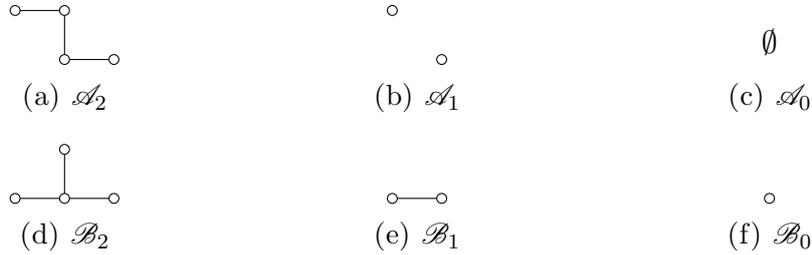

In this section we analyze the dynamic version of the model of Rydberg atoms on the ladder. Our main result is the closed formula for the jamming limit for all values of $b \ge 1$ (see \eqref{eq:expr_for_rho_b}). We denote the value of the jamming limit, when the blockade range is equal to $b$, by $\rho_{\infty}^b$. Plugging $b = 1$ in formula \eqref{eq:expr_for_rho_b} gives us
\begin{equation*}
    \rho_{\infty}^1 = \frac{1}{2} e^{-1} \left( \frac{1}{2} + \int_0^1 e^y dy \right) = \frac{1}{2} e^{-1} \left( \frac{1}{2} + e - 1 \right) = \frac{1}{2} e^{-1} \left( e -\frac{1}{2} \right) = \frac{1}{2} - \frac{1}{4e},
\end{equation*}
which recovers the result obtained in \cite{Fan1992, Baram1992, Georgiou2009, Chern2015}. Our approach is based on the technique originally used in \cite{Fan1992}, and then also applied in \cite{Georgiou2009} and \cite{Chern2015}. As already mentioned, we are interested in the model of Rydberg atoms on the ladder as the one shown in Figure \ref{fig:ladder}. However, after one of the neutral atoms is excited to a Rydberg state, the blockage effect leaves us with a graph that differs from the original ladder when it comes to its boundary. This is the reason why we need to study subgraphs of the ladder of the shapes shown in Figure \ref{fig:Ak_Bk}.
Denote by $A_k$ the expected number of excited atoms on the graph $\mathscr{A}_k$, and by $B_k$ the expected number of excited atoms on the graph $\mathscr{B}_k$. After one of the neutral atoms is excited to a Rydberg state, it causes blockage, see Figure \ref{fig:excitation}. In this, and all the following figures, bullets ($\f$) represent atoms excited to a Rydberg state, while empty bullets ($\e$) represent neutral atoms. Due to the blockage caused by the excited atom, the original graph is decomposed into two subgraphs that are again of the same shape as graphs introduced in Figure \ref{fig:Ak_Bk}. If the excited atom is close enough to the boundary, the original graph will not necessarily decompose into two graphs, but for simplicity of notation, we set $A_k = B_k = 0$ for every $k < 0$. Before proceeding, let us just clarify that $A_0 = 0$, but $B_0 = 1$ (see Figure \ref{fig:k=012}).

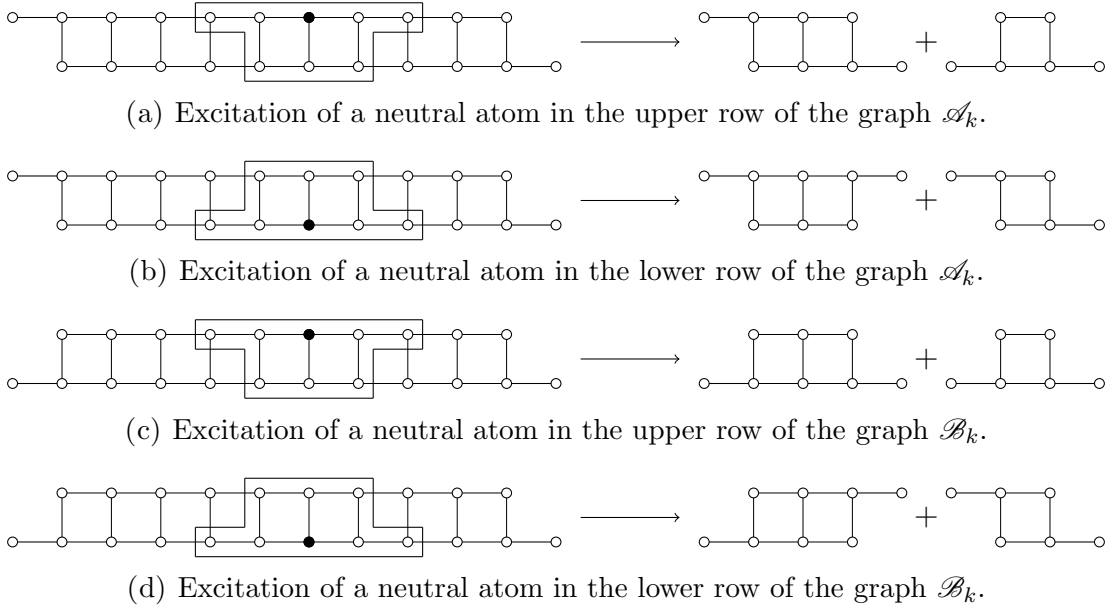
\begin{figure}
\begin{subfigure}{1.0\textwidth}
    \centering
    \begin{tikzpicture}[scale = 0.65]
		\draw[step=1cm,black,very thin] (0, 0) grid (9,1);
		\draw (0, 1) -- (-1, 1);
		\draw (9, 0) -- (10, 0);
		\foreach \x in {0,...,9}{
			\draw[color=black, fill=white] (\x,0) circle [radius=0.1];
			\draw[color=black, fill=white] (\x,1) circle [radius=0.1];}
            \draw[color=black, fill=white] (-1,1) circle [radius=0.1];
            \draw[color=black, fill=white] (10,0) circle [radius=0.1];
            \draw[color=black, fill=black] (5,1) circle [radius=0.1];
            \draw (2.7, 1.3) -- (7.3, 1.3);
            \draw (7.3, 1.3) -- (7.3, 0.7);
            \draw (7.3, 0.7) -- (6.3, 0.7);
            \draw (6.3, 0.7) -- (6.3, -0.3);
            \draw (6.3, -0.3) -- (3.7, -0.3);
            \draw (3.7, -0.3) -- (3.7, 0.7);
            \draw (3.7, 0.7) -- (2.7, 0.7);
            \draw (2.7, 0.7) -- (2.7, 1.3);

            \draw[->] (10.5, 0.5) -- (12.5, 0.5);

            \draw[step=1cm,black,very thin] (14, 0) grid (16,1);
		\draw (13, 1) -- (14, 1);
		\draw (16, 0) -- (17, 0);
		\foreach \x in {14,...,16}{
			\draw[color=black, fill=white] (\x,0) circle [radius=0.1];
			\draw[color=black, fill=white] (\x,1) circle [radius=0.1];}
            \draw[color=black, fill=white] (13,1) circle [radius=0.1];
            \draw[color=black, fill=white] (17,0) circle [radius=0.1];

            \node[rectangle, draw=none, minimum size=1pt] at (17.5, 0.5) {$+$};

            \draw[step=1cm,black,very thin] (19, 0) grid (20,1);
		\draw (18, 0) -- (19, 0);
		\draw (20, 0) -- (21, 0);
		\foreach \x in {19,...,20}{
			\draw[color=black, fill=white] (\x,0) circle [radius=0.1];
			\draw[color=black, fill=white] (\x,1) circle [radius=0.1];}
            \draw[color=black, fill=white] (18,0) circle [radius=0.1];
            \draw[color=black, fill=white] (21,0) circle [radius=0.1];
    \end{tikzpicture}
    \caption{Excitation of a neutral atom in the upper row of the graph $\mathscr{A}_k$.}
    \label{fig:Ak_upper_row}
\end{subfigure}

\bigskip

\begin{subfigure}{1.0\textwidth}
    \centering
    \begin{tikzpicture}[scale = 0.65]
		\draw[step=1cm,black,very thin] (0, 0) grid (9,1);
		\draw (0, 1) -- (-1, 1);
		\draw (9, 0) -- (10, 0);
		\foreach \x in {0,...,9}{
			\draw[color=black, fill=white] (\x,0) circle [radius=0.1];
			\draw[color=black, fill=white] (\x,1) circle [radius=0.1];}
            \draw[color=black, fill=white] (-1,1) circle [radius=0.1];
            \draw[color=black, fill=white] (10,0) circle [radius=0.1];
            \draw[color=black, fill=black] (5,0) circle [radius=0.1];
            \draw (2.7, -0.3) -- (7.3, -0.3);
            \draw (7.3, -0.3) -- (7.3, 0.3);
            \draw (7.3, 0.3) -- (6.3, 0.3);
            \draw (6.3, 0.3) -- (6.3, 1.3);
            \draw (6.3, 1.3) -- (3.7, 1.3);
            \draw (3.7, 1.3) -- (3.7, 0.3);
            \draw (3.7, 0.3) -- (2.7, 0.3);
            \draw (2.7, 0.3) -- (2.7, -0.3);

            \draw[->] (10.5, 0.5) -- (12.5, 0.5);

            \draw[step=1cm,black,very thin] (14, 0) grid (16,1);
		\draw (13, 1) -- (14, 1);
		\draw (16, 1) -- (17, 1);
		\foreach \x in {14,...,16}{
			\draw[color=black, fill=white] (\x,0) circle [radius=0.1];
			\draw[color=black, fill=white] (\x,1) circle [radius=0.1];}
            \draw[color=black, fill=white] (13,1) circle [radius=0.1];
            \draw[color=black, fill=white] (17,1) circle [radius=0.1];

            \node[rectangle, draw=none, minimum size=1pt] at (17.5, 0.5) {$+$};

            \draw[step=1cm,black,very thin] (19, 0) grid (20,1);
		\draw (18, 1) -- (19, 1);
		\draw (20, 0) -- (21, 0);
		\foreach \x in {19,...,20}{
			\draw[color=black, fill=white] (\x,0) circle [radius=0.1];
			\draw[color=black, fill=white] (\x,1) circle [radius=0.1];}
            \draw[color=black, fill=white] (18,1) circle [radius=0.1];
            \draw[color=black, fill=white] (21,0) circle [radius=0.1];
    \end{tikzpicture}
    \caption{Excitation of a neutral atom in the lower row of the graph $\mathscr{A}_k$.}
    \label{fig:Ak_lower_row}
\end{subfigure}

\bigskip

\begin{subfigure}{1.0\textwidth}
    \centering
    \begin{tikzpicture}[scale = 0.65]
		\draw[step=1cm,black,very thin] (0, 0) grid (9,1);
		\draw (0, 0) -- (-1, 0);
		\draw (9, 0) -- (10, 0);
		\foreach \x in {0,...,9}{
			\draw[color=black, fill=white] (\x,0) circle [radius=0.1];
			\draw[color=black, fill=white] (\x,1) circle [radius=0.1];}
            \draw[color=black, fill=white] (-1,0) circle [radius=0.1];
            \draw[color=black, fill=white] (10,0) circle [radius=0.1];
            \draw[color=black, fill=black] (5,1) circle [radius=0.1];
            \draw (2.7, 1.3) -- (7.3, 1.3);
            \draw (7.3, 1.3) -- (7.3, 0.7);
            \draw (7.3, 0.7) -- (6.3, 0.7);
            \draw (6.3, 0.7) -- (6.3, -0.3);
            \draw (6.3, -0.3) -- (3.7, -0.3);
            \draw (3.7, -0.3) -- (3.7, 0.7);
            \draw (3.7, 0.7) -- (2.7, 0.7);
            \draw (2.7, 0.7) -- (2.7, 1.3);

            \draw[->] (10.5, 0.5) -- (12.5, 0.5);

            \draw[step=1cm,black,very thin] (14, 0) grid (16,1);
		\draw (13, 0) -- (14, 0);
		\draw (16, 0) -- (17, 0);
		\foreach \x in {14,...,16}{
			\draw[color=black, fill=white] (\x,0) circle [radius=0.1];
			\draw[color=black, fill=white] (\x,1) circle [radius=0.1];}
            \draw[color=black, fill=white] (13,0) circle [radius=0.1];
            \draw[color=black, fill=white] (17,0) circle [radius=0.1];

            \node[rectangle, draw=none, minimum size=1pt] at (17.5, 0.5) {$+$};

            \draw[step=1cm,black,very thin] (19, 0) grid (20,1);
		\draw (18, 0) -- (19, 0);
		\draw (20, 0) -- (21, 0);
		\foreach \x in {19,...,20}{
			\draw[color=black, fill=white] (\x,0) circle [radius=0.1];
			\draw[color=black, fill=white] (\x,1) circle [radius=0.1];}
            \draw[color=black, fill=white] (18,0) circle [radius=0.1];
            \draw[color=black, fill=white] (21,0) circle [radius=0.1];
    \end{tikzpicture}
    \caption{Excitation of a neutral atom in the upper row of the graph $\mathscr{B}_k$.}
    \label{fig:Bk_upper_row}
\end{subfigure}

\bigskip

\begin{subfigure}{1.0\textwidth}
    \centering
    \begin{tikzpicture}[scale = 0.65]
		\draw[step=1cm,black,very thin] (0, 0) grid (9,1);
		\draw (0, 0) -- (-1, 0);
		\draw (9, 0) -- (10, 0);
		\foreach \x in {0,...,9}{
			\draw[color=black, fill=white] (\x,0) circle [radius=0.1];
			\draw[color=black, fill=white] (\x,1) circle [radius=0.1];}
            \draw[color=black, fill=white] (-1,0) circle [radius=0.1];
            \draw[color=black, fill=white] (10,0) circle [radius=0.1];
            \draw[color=black, fill=black] (5,0) circle [radius=0.1];
            \draw (2.7, -0.3) -- (7.3, -0.3);
            \draw (7.3, -0.3) -- (7.3, 0.3);
            \draw (7.3, 0.3) -- (6.3, 0.3);
            \draw (6.3, 0.3) -- (6.3, 1.3);
            \draw (6.3, 1.3) -- (3.7, 1.3);
            \draw (3.7, 1.3) -- (3.7, 0.3);
            \draw (3.7, 0.3) -- (2.7, 0.3);
            \draw (2.7, 0.3) -- (2.7, -0.3);

            \draw[->] (10.5, 0.5) -- (12.5, 0.5);

            \draw[step=1cm,black,very thin] (14, 0) grid (16,1);
		\draw (13, 0) -- (14, 0);
		\draw (16, 1) -- (17, 1);
		\foreach \x in {14,...,16}{
			\draw[color=black, fill=white] (\x,0) circle [radius=0.1];
			\draw[color=black, fill=white] (\x,1) circle [radius=0.1];}
            \draw[color=black, fill=white] (13,0) circle [radius=0.1];
            \draw[color=black, fill=white] (17,1) circle [radius=0.1];

            \node[rectangle, draw=none, minimum size=1pt] at (17.5, 0.5) {$+$};

            \draw[step=1cm,black,very thin] (19, 0) grid (20,1);
		\draw (18, 1) -- (19, 1);
		\draw (20, 0) -- (21, 0);
		\foreach \x in {19,...,20}{
			\draw[color=black, fill=white] (\x,0) circle [radius=0.1];
			\draw[color=black, fill=white] (\x,1) circle [radius=0.1];}
            \draw[color=black, fill=white] (18,1) circle [radius=0.1];
            \draw[color=black, fill=white] (21,0) circle [radius=0.1];
    \end{tikzpicture}
    \caption{Excitation of a neutral atom in the lower row of the graph $\mathscr{B}_k$.}
    \label{fig:Bk_lower_row}
\end{subfigure}
\caption{Illustration of all the possible scenarios of exciting a neutral atom to a Rydberg state (with blockade range $b$). After an atom becomes excited, due to the blockade range, the original graph decomposes into two subgraphs that are again of the two types introduced in Figure \ref{fig:Ak_Bk}. In this illustration, we have $k = 11$ and $b = 2$.}
\label{fig:excitation}
\end{figure}

The graph $\mathscr{A}_k$ (shown in Figure \ref{fig:Ak}) contains $2k$ sites occupied by neutral atoms. If an atom in the upper row of the graph $\mathscr{A}_k$ is excited, we are left with a graph of type $\mathscr{A}$ on the left hand side, and a graph of type $\mathscr{B}$ on the right hand side (see Figure \ref{fig:Ak_upper_row}). If, instead, the excited atom is one of the atoms in the lower row, after taking the blockage effect into account, we are left with a graph of type $\mathscr{B}$ on the left hand side, and a graph of type $\mathscr{A}$ on the right hand side (see Figure \ref{fig:Ak_lower_row}). Hence, it holds that
\begin{equation}\label{eq:rec_for_Ak}
    \begin{aligned}
        A_k
        & = 1 + \frac{1}{2k} \sum_{j = 0}^{k-1}(A_{j-b} + B_{k-j-b}) + \frac{1}{2k} \sum_{j = 1}^k(B_{j-b} + A_{k-j-b}) \\
        & = 1 + \frac{1}{k} \sum_{j = 1}^{k-1}(A_{j-b} + B_{j-b}) + \frac{1}{k} B_{k-b}.
    \end{aligned}
\end{equation}
Analyzing all the possible scenarios when starting from the graph $\mathscr{B}_k$ (shown in Figure \ref{fig:Bk}), in an analogous way (see Figures \ref{fig:Bk_upper_row} and \ref{fig:Bk_lower_row}) yields
\begin{equation}\label{eq:rec_for_Bk}
    \begin{aligned}
        B_k
        & = 1 + \frac{1}{2k} \sum_{j = 0}^k(A_{j-b} + A_{k-j-b}) + \frac{1}{2k} \sum_{j = 1}^{k-1}(B_{j-b} + B_{k-j-b})\\
        & = 1 + \frac{1}{k} \sum_{j = 1}^{k-1}(A_{j-b} + B_{j-b}) + \frac{1}{k} A_{k-b}.
    \end{aligned}
\end{equation}
We set $C_k = (A_k + B_k)/2$. In the thermodynamic limit $k\to\infty$ the values $A_k$, $B_k$, and $C_k$ are all identical. Summing equations \eqref{eq:rec_for_Ak} and \eqref{eq:rec_for_Bk} and dividing by $2$ gives
\begin{equation}\label{eq:rec_for_Ck}
    C_k = 1 + \frac{2}{k} \sum_{j = 1}^{k-1}C_{j-b} + \frac{1}{k}C_{k-b} = 1 + \frac{2}{k} \sum_{j = 0}^{k-b-1}C_j + \frac{1}{k}C_{k-b},
\end{equation}
where $C_k = 0$, for every $k < 0$, and $C_0 = (A_0 + B_0)/2 = 1/2$ (see Figure \ref{fig:k=012}). By multiplying \eqref{eq:rec_for_Ck} through by $kx^{k-1}$ and summing over all $k$ from one to infinity we get
\begin{equation}\label{eq:de_for_C_1st_part}
    \begin{aligned}
        \sum_{k = 1}^{\infty} C_k k x^{k-1}
        & = \sum_{k = 1}^{\infty} k x^{k-1} + 2\sum_{j = 0}^{\infty}C_j \sum_{k = j+b+1}^{\infty}x^{k-1} + \sum_{k = 1}^{\infty} C_{k-b}x^{k-1} \\
        & = \frac{1}{(1-x)^2} + 2\sum_{j = 0}^{\infty}C_j \frac{x^{j+b}}{1-x} + x^{b-1}\sum_{k = 1}^{\infty}C_{k-b} x^{k-b}.
\end{aligned}
\end{equation}
Setting $C(x) = \sum_{k = 0}^{\infty}C_k x^k$, we immediately get from \eqref{eq:de_for_C_1st_part}
\begin{equation}\label{eq:de_for_C}
    C'(x) = \frac{1}{(1-x)^2} + \frac{2x^b}{1-x}C(x) + x^{b-1}C(x) = \frac{x^b + x^{b-1}}{1-x} C(x) + \frac{1}{(1-x)^2}.
\end{equation}
By using
\begin{equation*}
    \frac{x^b + x^{b-1}}{1-x} = -x^{b-1} - 2x^{b-2} - 2x^{b-3} - \cdots - 2x - 2 + \frac{2}{1-x},
\end{equation*}
one easily gets the solution of the homogeneous equation associated with \eqref{eq:de_for_C} as
\begin{equation}\label{eq:C_hom}
    C_{\text{hom}}(x) = D \exp \left( -\frac{x^b}{b} - 2\sum_{j=1}^{b-1}\frac{x^j}{j} \right) \cdot \frac{1}{(1-x)^2},
\end{equation}
for some constant $D\in \mathbb{R}$. Variation of the constant $D$ gives us
\begin{equation}\label{eq:de_for_D}
    D'(x) = \exp \left( \frac{x^b}{b} + 2\sum_{j = 1}^{b-1} \frac{x^j}{j} \right).
\end{equation}
By combining equations \eqref{eq:C_hom} and \eqref{eq:de_for_D}, and by taking into account $C(0) = C_0 = 1/2$, we get
\begin{equation}\label{eq:expr_for_C}
    C(x) = \frac{1}{(1-x)^2} \exp \left( -\frac{x^b}{b} - 2\sum_{j = 1}^{b-1} \frac{x^j}{j} \right) \left( \frac{1}{2} + \int_0^x \exp \left( \frac{y^b}{b} + 2\sum_{j = 1}^{b-1} \frac{y^j}{j} \right) dy \right).
\end{equation}
Now we are ready to calculate the jamming limit for an arbitrary value of the blockade range $b \ge 1$. We want to calculate $\rho_{\infty}^b = \lim_{k \to \infty} C_k/2k$. Combining the fact that $\lim_{k \to \infty} C_k/k = \lim_{x \to 1} (1 - x)^2 C(x)$, and the formula for $C(x)$ given in \eqref{eq:expr_for_C}, we get
\begin{equation}\label{eq:expr_for_rho_b}
    \begin{aligned}
        \rho_{\infty}^b
        & = \frac{1}{2} \exp \left( -\frac{1}{b} - 2\sum_{j = 1}^{b-1} \frac{1}{j} \right) \left( \frac{1}{2} + \int_0^1 \exp \left( \frac{y^b}{b} + 2\sum_{j = 1}^{b-1} \frac{y^j}{j} \right) dy \right) \\
        & = \frac{1}{2 H_b(1)} \left( \frac{1}{2} + \int_0^1 H_b(y) dy \right),
    \end{aligned}
\end{equation}
where
\begin{equation*}
    H_b(x) = \exp \left( \frac{x^b}{b} + 2\sum_{j = 1}^{b-1} \frac{x^j}{j} \right).
\end{equation*}

\section{Equilibrium model}\label{sec:equilibrium_model}
\begin{figure}
\centering
\begin{subfigure}{0.3\textwidth}
    \centering
    \begin{tikzpicture}[scale = 0.65]
	\draw[step=1cm,black,very thin] (0, 0) grid (6,1);
	\foreach \x in {0,...,6}{
            \draw[color=black, fill=white] (\x,0) circle [radius=0.1];
            \draw[color=black, fill=white] (\x,1) circle [radius=0.1];}
        \draw[color=black, fill=black] (0,0) circle [radius=0.1];
        \draw[color=black, fill=black] (2,1) circle [radius=0.1];
        \draw[color=black, fill=black] (4,0) circle [radius=0.1];
        \draw[color=black, fill=black] (6,1) circle [radius=0.1];
    \end{tikzpicture}
    \caption{$N = 4, L = 7$}
    \label{fig:N4L7}
\end{subfigure}
\begin{subfigure}{0.3\textwidth}
    \centering
    \begin{tikzpicture}[scale = 0.65]
	\draw[step=1cm,black,very thin] (0, 0) grid (6,1);
	\foreach \x in {0,...,6}{
            \draw[color=black, fill=white] (\x,0) circle [radius=0.1];
            \draw[color=black, fill=white] (\x,1) circle [radius=0.1];}
        \draw[color=black, fill=black] (1,1) circle [radius=0.1];
        \draw[color=black, fill=black] (4,1) circle [radius=0.1];
        \draw[color=black, fill=black] (6,0) circle [radius=0.1];
    \end{tikzpicture}
    \caption{$N = 3, L = 7$}
    \label{fig:N3L7}
\end{subfigure}
\begin{subfigure}{0.3\textwidth}
    \centering
    \begin{tikzpicture}[scale = 0.65]
	\draw[step=1cm,black,very thin] (0, 0) grid (6,1);
	\foreach \x in {0,...,6}{
            \draw[color=black, fill=white] (\x,0) circle [radius=0.1];
            \draw[color=black, fill=white] (\x,1) circle [radius=0.1];}
        \draw[color=black, fill=black] (1,0) circle [radius=0.1];
        \draw[color=black, fill=black] (5,1) circle [radius=0.1];
    \end{tikzpicture}
    \caption{$N = 2, L = 7$}
    \label{fig:N2L7}
\end{subfigure}
\caption{Three jammed configurations of Rydberg atoms on the ladder of length $L = 7$, with blockade range $b = 2$. The number of Rydberg atoms in these configurations is $N = 4, 3, 2$ (from left to right).}
\label{fig:JC_on_L=7}
\end{figure}
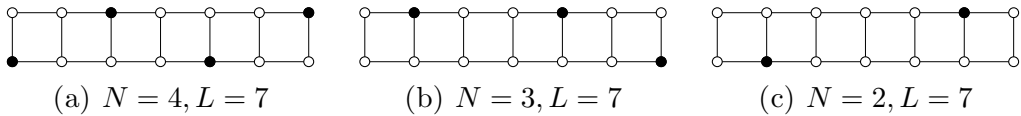

The main goal of this section is to obtain the complexity function for the equilibrium model of Rydberg atoms on the ladder for all $b \ge 1$. Recently, a renewal approach for calculating the complexity function of a model was introduced in \cite{Krapivsky2023} (see also \cite{Luck2024}). Inspired by the technique used there, we aim to compute a bivariate generating function enumerating all the jammed configurations on the ladder with $2L$ sites (hence, of length $L$) with precisely $N$ atoms excited to a Rydberg state (when the blockade range is equal to $b$). Before explaining the procedure, let us inspect some concrete examples of jammed configurations to get a better feeling about their possible shapes. Figure \ref{fig:JC_on_L=7} displays three different jammed configurations on the ladder of length $L = 7$, where the blockade range $b$ is equal to two, i.e., any two excited atoms are at least two sites apart.

We now show how to build all the jammed configurations from the parts of the ladder that only contain excited atoms in the upper row, and the parts of the ladder that only contain excited atoms in the lower row. Notice that two consecutive excited atoms in the upper row must be separated by at least $b$ neutral atoms (so that the constraint imposed by the blockage effect is satisfied), and at most $2b - 2$ neutral atoms (since otherwise we would not end up with a jammed configuration, see Figure \ref{fig:max_violation}). Hence, the part of the ladder that only contains excited atoms in the upper row will be composed of blocks that have the first atom in the upper row excited to a Rydberg state, followed by a cluster of at least $2b+1$, and at most $2(2b - 2)+1$, neutral atoms (i.e.\ after the first column where we have an excited atom in the upper row, we have between $b$ and $2b - 2$ empty columns, see Figure \ref{fig:building_blocks_of_Fb}).
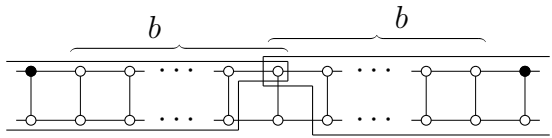
\begin{figure}
    \centering
    \begin{tikzpicture}[scale = 0.65]
        \draw[step=1cm,black,very thin] (0, 0) grid (2,1);
		\draw (0, 0) -- (-0.3, 0);
		\draw (0, 1) -- (-0.3, 1);
            \draw (2, 0) -- (2.3, 0);
		\draw (2, 1) -- (2.3, 1);
		\foreach \x in {0,...,2}{
			\draw[color=black, fill=white] (\x,0) circle [radius=0.1];
			\draw[color=black, fill=white] (\x,1) circle [radius=0.1];}
            \draw[color=black, fill=black] (0,1) circle [radius=0.1];
            \node[rectangle, draw=none, minimum size=1pt] at (3, 0) {$\cdots$};
            \node[rectangle, draw=none, minimum size=1pt] at (3, 1) {$\cdots$};
            \draw[step=1cm,black,very thin] (4, 0) grid (6,1);
		\draw (4, 0) -- (3.7, 0);
		\draw (4, 1) -- (3.7, 1);
            \draw (6, 0) -- (6.3, 0);
            \draw (6, 1) -- (6.3, 1);
		\foreach \x in {4,...,6}{
			\draw[color=black, fill=white] (\x,0) circle [radius=0.1];
			\draw[color=black, fill=white] (\x,1) circle [radius=0.1];}
            \node[rectangle, draw=none, minimum size=1pt] at (7, 0) {$\cdots$};
            \node[rectangle, draw=none, minimum size=1pt] at (7, 1) {$\cdots$};
            \draw[step=1cm,black,very thin] (8, 0) grid (10,1);
		\draw (8, 0) -- (7.7, 0);
		\draw (8, 1) -- (7.7, 1);
            \draw (10, 0) -- (10.3, 0);
            \draw (10, 1) -- (10.3, 1);
		\foreach \x in {8,...,10}{
			\draw[color=black, fill=white] (\x,0) circle [radius=0.1];
			\draw[color=black, fill=white] (\x,1) circle [radius=0.1];}
            \draw[color=black, fill=black] (10,1) circle [radius=0.1];
            \draw [decorate, decoration = {calligraphic brace}] (.8,1.4) -- (5.2,1.4);
            \draw [decorate, decoration = {calligraphic brace}] (4.8,1.6) -- (9.2,1.6);
            \draw (-0.5, 1.2) -- (5.2, 1.2);
            \draw (5.2, 1.2) -- (5.2, 0.8);
            \draw (5.2, 0.8) -- (4.2, 0.8);
            \draw (4.2, 0.8) -- (4.2, -0.2);
            \draw (4.2, -0.2) -- (-0.5, -0.2);
            \draw (4.7, 1.3) -- (10.5, 1.3);
            \draw (4.7, 1.3) -- (4.7, 0.7);
            \draw (4.7, 0.7) -- (5.7, 0.7);
            \draw (5.7, -0.3) -- (5.7, 0.7);
            \draw (5.7, -0.3) -- (10.5, -0.3);
            \node[rectangle, draw=none, minimum size=1pt] at (2.5, 1.9) {$b$};
            \node[rectangle, draw=none, minimum size=1pt] at (7.5, 2.1) {$b$};
    \end{tikzpicture}
    \caption{If two consecutive excited atoms are both in the upper row, and are separated by $2b - 1$ (or more) neutral atoms, we will not have a jammed configuration, because there will be at least one site in the lower row (between those two excited atoms in the upper row) where we will be allowed to excite an atom to a Rydberg state.}
    \label{fig:max_violation}
\end{figure}
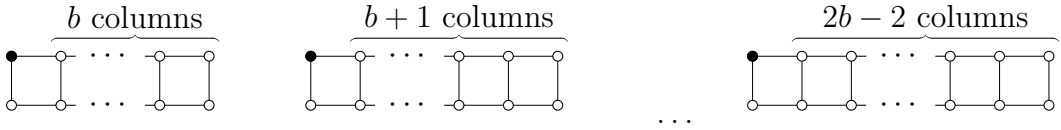
\begin{figure}
\centering
\begin{subfigure}{0.25\textwidth}
    \centering
    \begin{tikzpicture}[scale = 0.65]
	\draw[step=1cm,black,very thin] (0, 0) grid (1,1);
        \draw (1, 0) -- (1.3, 0);
        \draw (1, 1) -- (1.3, 1);
	\foreach \x in {0,...,1}{
            \draw[color=black, fill=white] (\x,0) circle [radius=0.1];
            \draw[color=black, fill=white] (\x,1) circle [radius=0.1];}
        \draw[color=black, fill=black] (0,1) circle [radius=0.1];
        \node[rectangle, draw=none, minimum size=1pt] at (2, 0) {$\cdots$};
        \node[rectangle, draw=none, minimum size=1pt] at (2, 1) {$\cdots$};
        \draw[step=1cm,black,very thin] (3, 0) grid (4,1);
        \draw (2.7, 0) -- (3, 0);
        \draw (2.7, 1) -- (3, 1);
	\foreach \x in {3,...,4}{
            \draw[color=black, fill=white] (\x,0) circle [radius=0.1];
            \draw[color=black, fill=white] (\x,1) circle [radius=0.1];}
        \draw [decorate, decoration = {calligraphic brace}] (0.8, 1.3) -- (4.2, 1.3);
        \node[rectangle, draw=none, minimum size=1pt] at (2.5, 1.8) {$b$ columns};
    \end{tikzpicture}
\end{subfigure}
\begin{subfigure}{0.3\textwidth}
    \centering
    \begin{tikzpicture}[scale = 0.65]
	\draw[step=1cm,black,very thin] (0, 0) grid (1,1);
        \draw (1, 0) -- (1.3, 0);
        \draw (1, 1) -- (1.3, 1);
	\foreach \x in {0,...,1}{
            \draw[color=black, fill=white] (\x,0) circle [radius=0.1];
            \draw[color=black, fill=white] (\x,1) circle [radius=0.1];}
        \draw[color=black, fill=black] (0,1) circle [radius=0.1];
        \node[rectangle, draw=none, minimum size=1pt] at (2, 0) {$\cdots$};
        \node[rectangle, draw=none, minimum size=1pt] at (2, 1) {$\cdots$};
        \draw[step=1cm,black,very thin] (3, 0) grid (5,1);
        \draw (2.7, 0) -- (3, 0);
        \draw (2.7, 1) -- (3, 1);
	\foreach \x in {3,...,5}{
            \draw[color=black, fill=white] (\x,0) circle [radius=0.1];
            \draw[color=black, fill=white] (\x,1) circle [radius=0.1];}
        \draw [decorate, decoration = {calligraphic brace}] (0.8, 1.3) -- (5.2, 1.3);
        \node[rectangle, draw=none, minimum size=1pt] at (3, 1.8) {$b+1$ columns};
    \end{tikzpicture}
\end{subfigure}
\begin{subfigure}{0.1\textwidth}
$$\centering\ldots$$
\end{subfigure}
\begin{subfigure}{0.3\textwidth}
    \centering
    \begin{tikzpicture}[scale = 0.65]
	\draw[step=1cm,black,very thin] (0, 0) grid (2,1);
        \draw (2, 0) -- (2.3, 0);
        \draw (2, 1) -- (2.3, 1);
	\foreach \x in {0,...,2}{
            \draw[color=black, fill=white] (\x,0) circle [radius=0.1];
            \draw[color=black, fill=white] (\x,1) circle [radius=0.1];}
        \draw[color=black, fill=black] (0,1) circle [radius=0.1];
        \node[rectangle, draw=none, minimum size=1pt] at (3, 0) {$\cdots$};
        \node[rectangle, draw=none, minimum size=1pt] at (3, 1) {$\cdots$};
        \draw[step=1cm,black,very thin] (4, 0) grid (6,1);
        \draw (3.7, 0) -- (4, 0);
        \draw (3.7, 1) -- (4, 1);
	\foreach \x in {4,...,6}{
            \draw[color=black, fill=white] (\x,0) circle [radius=0.1];
            \draw[color=black, fill=white] (\x,1) circle [radius=0.1];}
        \draw [decorate, decoration = {calligraphic brace}] (0.8, 1.3) -- (6.2, 1.3);
        \node[rectangle, draw=none, minimum size=1pt] at (3.5, 1.8) {$2b-2$ columns};
    \end{tikzpicture}
\end{subfigure}
\caption{Building blocks of the parts of jammed configurations where excited atoms are only in the upper row.}
\label{fig:building_blocks_of_Fb}
\end{figure}
These building blocks are encoded by the polynomial
\begin{equation}\label{eq:def_of_Pb}
    P_b(x, y) = xy^{2(b+1)} + \ldots + xy^{2(2b-1)},
\end{equation}
where $x$ is a formal variable associated with the number of atoms excited to a Rydberg state, and $y$ is a formal variable associated with the total number of atoms (neutral and excited) in the ladder of fixed length. Notice that any block from Figure \ref{fig:building_blocks_of_Fb} can be glued to any other block from that figure. After the part of the ladder where we only have excited atoms in the upper row, we switch to the part where we only have atoms in the lower row. To be able to make a transition from the part where excited atoms are only in the upper row, to the part where excited atoms are only in the lower row, the former has to end with at least $b-1$ empty columns (and then an excited atom can appear in the lower row without violating the blockage constraint), and at most $2b-1$ empty columns (after which we must have an excited atom in the lower row to keep the configuration jammed). This last block in the part of the jammed configuration that only contains excited atoms in the upper row is, therefore, encoded by the polynomial
\begin{equation}\label{eq:def_of_Eb}
     E_b(x, y) = xy^{2b} + \ldots + xy^{2\cdot 2b}.
\end{equation}
Combining \eqref{eq:def_of_Pb} and \eqref{eq:def_of_Eb}, we get that the bivariate generating function associated with the part that only contains excited atoms in the upper row is given by
\begin{equation}\label{eq:def_of_Fb}
     F_b(x, y) = \sum_{n = 0}^{\infty} (P_b(x, y))^n \cdot E_b(x, y) = \frac{xy^{2b} (1 - y^{2(b+1)})}{1 - y^2 - xy^{2(b+1)} (1 - y^{2(b-1)})}.
\end{equation}
Clearly, due to the symmetry between the upper and the lower row, the situation is completely analogous when we consider parts of the jammed configuration that only have excited atoms in the lower row. Hence, by using function $F_b(x, y)$ from \eqref{eq:def_of_Fb}, we can encode the whole jammed configuration, except its beginning and end. Let us now inspect what can happen at the beginning and at the end of a jammed configuration. Since blocks in Figure \ref{fig:building_blocks_of_Fb} all start with an excited atom, we can have some empty columns before the first such block. These empty columns will clearly not violate the blockage effect, but to keep our configuration jammed, the number of empty columns at the beginning can be between $0$ and $b-1$. These empty columns at the beginning of a jammed configuration are encoded by the polynomial
\begin{equation*}
    S_b(x, y) = 1 + y^2 + \cdots + y^{2(b-1)}.
\end{equation*}
The end of the configuration can again be interpreted as one of the parts that has excited atoms only in the upper/lower row. The only difference is that the last part has to end in a way to keep the configuration jammed. This means that, using the same polynomial $P_b(x, y)$ introduced in \eqref{eq:def_of_Pb}, and only changing the end polynomial $E_b(x, y)$ defined in \eqref{eq:def_of_Eb} in an appropriate way, will give us the generating function for the last part of the jammed configuration. After the last excited atom (just like before the first one), we can have from $0$ to $b-1$ empty columns. Therefore, we will end the last part of the jammed configuration with a block that has excited atom in the first column, and then between $0$ and $b-1$ empty columns. We encode this by the polynomial
\begin{equation}\label{eq:def_of_Eb_tilde}
     \widetilde{E}_b(x, y) = xy^2 + xy^4 + \ldots + xy^{2b}.
\end{equation}
This gives us
\begin{equation*}
    \widetilde{F}_b = \sum_{n = 0}^{\infty} (P_b(x, y))^n \cdot \widetilde{E}_b(x, y) = \frac{xy^2 (1 - y^{2b})}{1 - y^2 - xy^{2(b+1)} (1 - y^{2(b-1)})}.
\end{equation*}
Now we have all the ingredients for calculating the bivariate generating function enumerating all the jammed configurations on the ladder with $2L$ sites (hence, of length $L$) with precisely $N$ atoms excited to a Rydberg state (when the blockade range is equal to $b$). We denote this bivariate generating function by $G_b(x, y)$:
\begin{align}\label{eq:G}
   	G_b(x, y) =& 1 + S_b(x, y) \cdot 2 \sum_{n = 0}^{\infty} (F_b(x, y))^n \cdot \widetilde{F}_b(x, y) = 1 + \frac{2S_b(x, y)\widetilde{F}_b(x, y)}{1 - F_b(x, y)}\nonumber\\
 =& 1 + \frac{2xy^2(1-y^{2b})^2}{(1-y^2)^2-xy^{2b}(1-y^{2b})(1-y^4)},
\end{align}
where the factor $2$ appears since the leftmost excited atom can appear in the upper, or in the lower row. As an example, expanding the function $G_b(x, y)$ into the Taylor series at $y=0$, when $b=2$, the term containing $y^{14}$ is
$$(2 x^2 + 30 x^3 + 2 x^4)y^{14}$$
from which one can read that there are in total $34$ jammed configurations on the ladder of length $L = 7$, $30$ of which have $3$ excited atoms, $2$ have $2$ excited atoms, and the remaining $2$ have $4$ excited atoms. Notice that the jammed configuration shown in Figure \ref{fig:N4L7}, and its symmetric version, where the first atom from the left is in the upper row, are the only ones with $4$ excited atoms. Similarly, the jammed configuration in Figure \ref{fig:N2L7}, and its symmetric version, are the only ones with $2$ excited atoms.

\begin{figure}
	\begin{subfigure}{.5\textwidth}\centering
		\includegraphics[width=\textwidth]{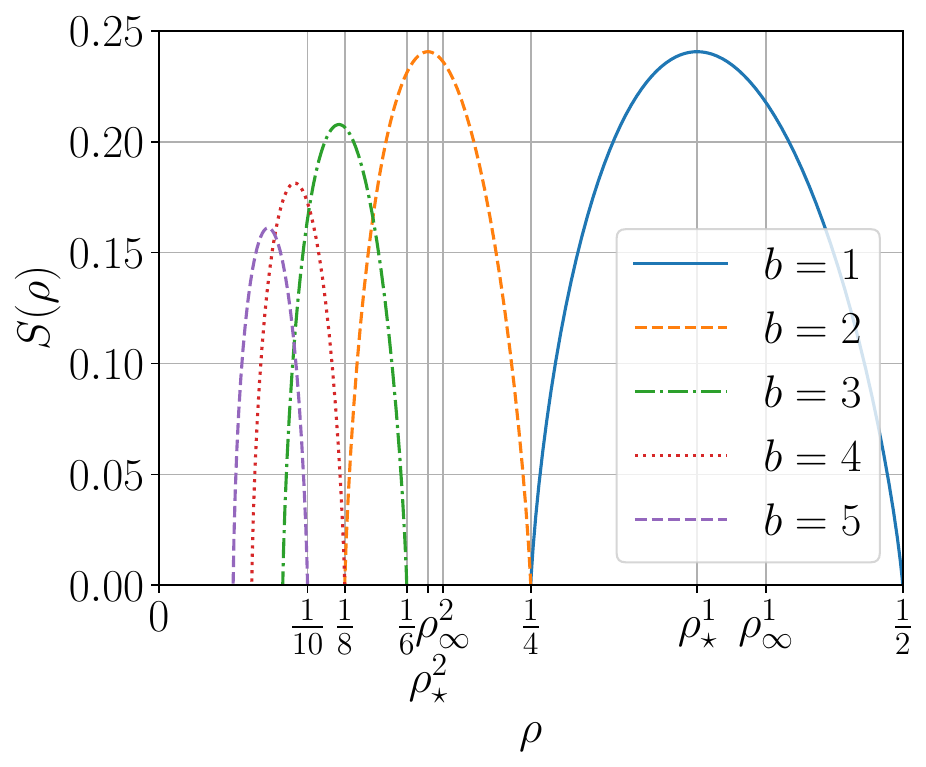}
		\caption{}\label{fig:S_b_1}
	\end{subfigure}%
	\begin{subfigure}{.5\textwidth}\centering
		\includegraphics[width=\textwidth]{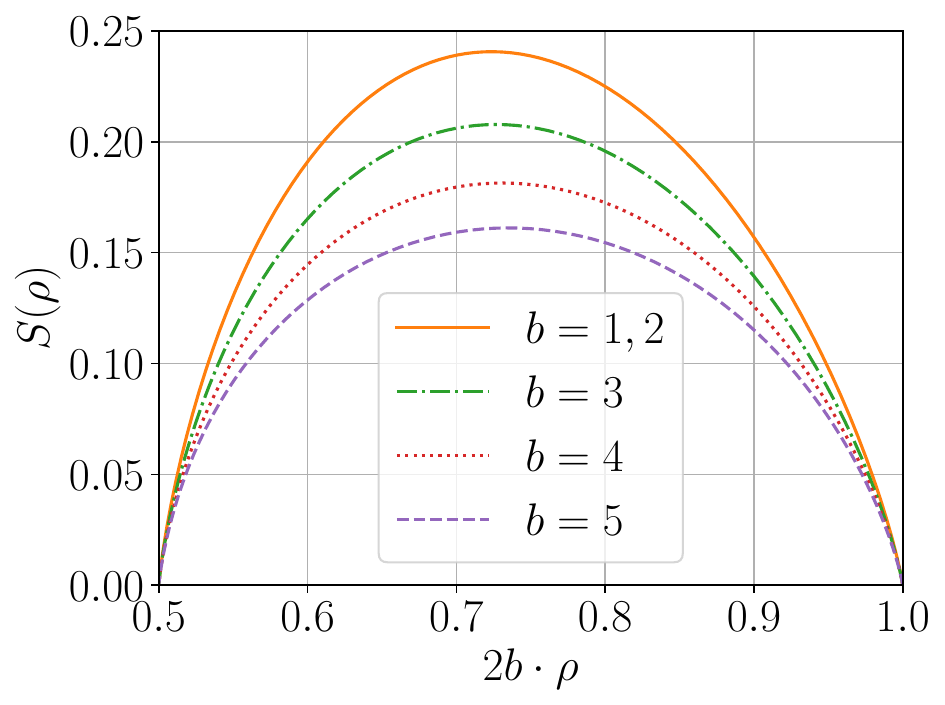}
		\caption{}\label{fig:S_b_2}
	\end{subfigure}
	\caption{Complexity function of the Rydberg atoms on the ladder model for $1 \le b \le 5$ (left) and the same graphs scaled by $2b$ in the horizontal direction in order to have the same support $[0.5,1]$. Note that the rescaled graphs for $b=1$ and $b=2$ are identical.}\label{fig:S_b}
\end{figure}

We now proceed to compute the complexity function for this model by using the method from \cite{Krapivsky2023} (see also \cite{Puljiz2023b} for details). For this, we need to extract the denominator of the rational function $G_b(x, y)$. It is straightforward to see that a factor of $(1-y^2)^2$ can be factored out from both the numerator and denominator in \eqref{eq:G}, which means that the denominator of $G_b(x, y)$ can be taken to be
\begin{equation}\label{eq:def_of_q}
	\begin{split}
	    q_b(x, y) &= \frac{(1-y^2)^2-xy^{2b}(1-y^{2b})(1-y^4)}{(1-y^2)^2} = 1 - xy^{2b}\frac{1 - y^{2b}}{1-y^2}(1 + y^2)\\
	    &= 1-xy^{2b}(1+2y^2+2y^4+\dots+2y^{2b-2}+y^{2b}).
	\end{split}
\end{equation}
Now for any $x_0>0$, $y_0>0$ is taken to be the unique positive solution to the equation $q_b(x_0,y_0)=0$, and therefore
\begin{equation}\label{eq:x_in_terms_of_y}
	x_0 = \frac{1}{y_0^{2b}(1+2y_0^2+2y_0^4+\dots+2y_0^{2b-2}+y_0^{2b})}.
\end{equation}
The complexity function can now be expressed as
$$S(\rho) = -\rho \ln x_0 - \ln y_0,$$
where
\begin{equation}\label{eq:rho_via_y}
	\begin{aligned}
		\rho &= \left[\frac{x}{y} \frac{\partial_x q_b}{\partial_y q_b}\right]_{x=x_0,y=y_0} =\frac{(1-y_0^4)(1-y_0^{2b})}{4y_0^2(1-y_0^{2b})+2b(1-y_0^4)(1-2y_0^{2b})}\\
		&= \frac{1+2y_0^2+\dots+2y_0^{2b-2}+y_0^{2b}}{2b(1+2y_0^2+\dots+2y_0^{2b-2}+y_0^{2b})+2(2y_0^2+4y_0^4+\dots+(2b-2)y_0^{2b-2}+by_0^{2b})}.
	\end{aligned}
\end{equation}
The equations \eqref{eq:x_in_terms_of_y} and \eqref{eq:rho_via_y} parameterize the complexity function of the model $S(\rho)$ as a function of the parameter $y_0>0$. Figure \ref{fig:S_b_1} shows graphs of $S(\rho)$ for several different values of $b$. As the problem formulation suggests, scaling all the densities by $2b$ conveniently places them on the same support $[0.5,1]$. The resulting graphs are given in Figure \ref{fig:S_b_2}.
\begin{remark}
	It is not hard to calculate the explicit form of the complexity function in cases $b=1,2$:
	\begin{align*}
		S^1(\rho) &= \frac12\left[ 2\rho \ln(2\rho) - \left(1-2\rho\right) \ln(1-2\rho) - \left(4\rho-1\right) \ln(4\rho -1) \right], &\text{for } \frac14 < \rho < \frac12, \\
		S^2(\rho) &= \frac12\left[ 4\rho \ln(4\rho) - \left(1-4\rho\right) \ln(1-4\rho) - \left(8\rho-1\right) \ln(8\rho -1) \right], &\text{for } \frac18 < \rho < \frac14.
	\end{align*}
	Note that scaling the densities by $2$ in the first expression, and by $4$ in the second, produces the same function. This is visible in Figure \ref{fig:S_b_2}. We also note that this same function (modulo scaling) was already computed for an equivalent model by Krapivsky and Luck in \cite[equation (4.28)]{Krapivsky2023} and by the last three authors in \cite[equation (2.14)]{Puljiz2023b}.
\end{remark}

The equilibrium density $\rho_\star$ at which the complexity function attains its maximum is known to correspond to the value $x_0=1$. From \eqref{eq:x_in_terms_of_y} we see that the associated $0<y_\star<1$ solves
\begin{equation}\label{eq:ystar}
	y_\star^{2b}(1+2y_\star^2+2y_\star^4+\dots+2y_\star^{2b-2}+y_\star^{2b})=y_\star^{2b}\frac{1 - y_\star^{2b}}{1-y_\star^2}(1 + y_\star^2)=1
\end{equation}
and plugging this $y_\star$ into \eqref{eq:rho_via_y} gives the equilibrium density $\rho_\star^b$. We list scaled numerical values for a few of these constants alongside the scaled jamming limits, calculated in \eqref{eq:expr_for_rho_b}, in Table \ref{tab:rhos} (see also Figure \ref{fig:rhos_2b}).
\begin{table}
	\caption{Scaled jamming limits and equilibrium densities for a few values of $b$ (all the values are rounded to four decimal places).}\label{tab:rhos}
	\begin{tabular}{c||c|c|}
		$b$ & $2b\cdot\rho_\infty^b$ & $2b\cdot\rho_\star^b$\\
		\hline\hline
		$1$ & $1-(2e)^{-1} \approx 0.8161$ & $(5+\sqrt5)/10\approx0.7236$ \\
		$2$ & $e^{-5/2} + e^{-9/2}\sqrt{2\pi}\left(\operatorname{erfi}{\left(3/\sqrt2\right)} - \operatorname{erfi}{\left(\sqrt2\right)}\right) \approx 0.7634$ & $(5+\sqrt5)/10\approx0.7236$ \\
		$3$ & $0.7545$ & $0.7267$ \\
		$4$ & $0.7514$ & $0.7309$ \\
		$5$ & $0.7501$ & $0.7350$ \\
		$10$ & $0.7482$ & $0.7509$ \\
		$100$ & $0.7476$ & $0.8097$\\
		$\infty$ & $\int_0^\infty \exp\left(-2\int_0^{y} \frac{1-e^{-x}}{x} dx\right) dy \approx0.7476$ & $1$\\
		\hline
	\end{tabular}
\end{table}
In the next section we justify the values in the last row of Table \ref{tab:rhos} corresponding to the limiting values of the scaled jamming limit and equilibrium density as $b\to\infty$.

\section{Comparison of dynamic and equilibrium model}\label{sec:comparison}
\begin{figure}
	\includegraphics[width=.8\textwidth]{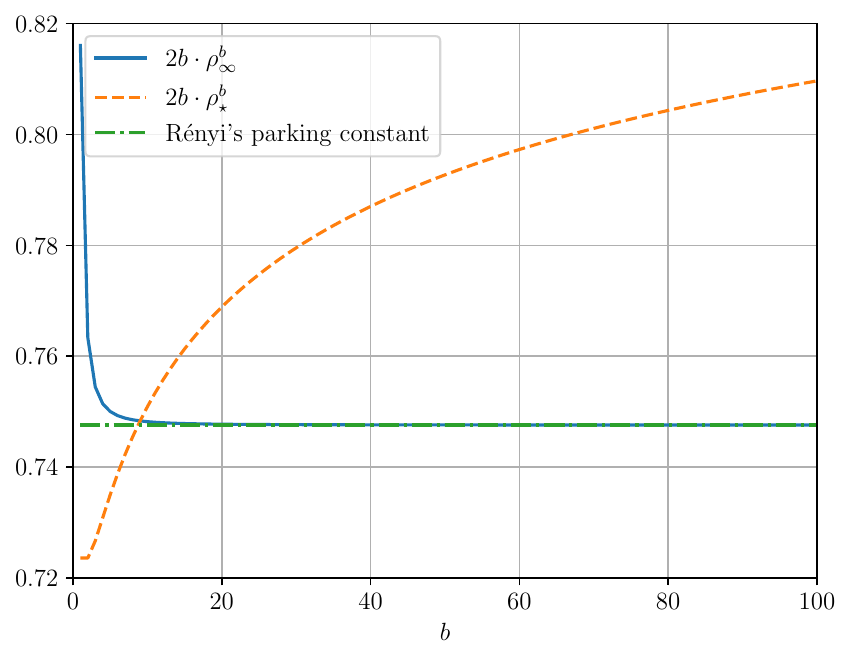}
	\caption{Scaled jamming limits and equilibrium densities for $1\le b\le 100$.}\label{fig:rhos_2b}
\end{figure}

In this section, we compare the models from Sections \ref{sec:dynamic_model} and \ref{sec:equilibrium_model}. The complexity function, computed in Section \ref{sec:equilibrium_model}, is displayed in Figure \ref{fig:S_b_1}, for $1 \le b \le 5$. The values $\rho_{\infty}^b$ and $\rho_{\star}^b$ are also indicated for $b=1,2$. It is evident that the Edwards hypothesis is violated. Let us now check what happens when $b$ tends to infinity. Clearly, both $\rho_{\infty}^b$ and $\rho_{\star}^b$ tend to zero as $b$ tends to infinity. As we have mentioned before, it is much more interesting (and natural) to inspect the behavior of sequences $(2b\cdot\rho_{\infty}^b)_b$ and $(2b\cdot\rho_{\star}^b)_b$. Table \ref{tab:rhos} contains these scaled densities and they are also depicted in Figure \ref{fig:rhos_2b}. In this way, not only do we count the excited atoms, but with each atom we also include the block of $2b$ atoms surrounding that atom ($b$ in each of the rows of the ladder). Intuitively, we are in the setting of parking cars on a line. The original (continuous) car parking problem was posed, and solved, by R\'{e}nyi in \cite{Renyi1958}. A discrete version of that problem was studied by Page in \cite{Page1959}. In the discrete version of R\'{e}nyi's car parking problem from \cite{Page1959}, cars of length $2$ were studied. The same problem was also studied in the context of irreversible deposition of $k$-mers on a linear substrate (see \cite{Gonzalez1974, Krapivsky2010a,Doslic2019}). The model from \cite{Page1959} corresponds to the deposition of dimers on a linear substrate. However, by increasing the length of a $k$-mer (i.e.\ the length of a car), and refining the linear substrate onto which those $k$-mers are deposited, we are getting better and better approximation of the continuous car-parking problem. It was shown in \cite{Gonzalez1974} that, as $k$ grows to infinity, the jamming limit from the model of irreversible deposition of $k$-mers on a linear substrate converges to the jamming limit of the continuous model. The jamming limit in the continuous model is the so-called R\'{e}nyi's car-parking constant which is equal to
\begin{equation}\label{eq:Renyis-cp-const}
    \int_0^\infty \exp\left(-2\int_0^{y} \frac{1-e^{-x}}{x} dx\right)dy = 0.7475979202\dots
\end{equation}

Clearly, there is a difference between the model for Rygberg atoms on a one-dimensional lattice, and on the ladder. This difference is demonstrated by the difference in the expressions for the jamming limits in the one-dimensional case, see \cite{Friedman1964, Mackenzie1962, Krapivsky2020}, and on the ladder, see \eqref{eq:expr_for_rho_b}. However, both models can be interpreted as a discrete approximation of the continuous car-parking problem. Hence, just as in the case of Rydberg atoms on the one-dimensional lattice (see \cite{Doslic2024a, Krapivsky2023}), we expect that the limit of the sequence $(2b \rho_{\infty}^b)_b$, as $b$ tends to infinity, is equal to R\'{e}nyi's car-parking constant. Let us now determine the limit $\lim_{b \to \infty}2b\rho_{\infty}^b$, and check whether we get the expected result. We start with rewriting the formula for $\rho_{\infty}^b$ from relation \eqref{eq:expr_for_rho_b}. We have
\begin{equation}\label{eq:expr_for_rho_b_1-y^j}
    \begin{aligned}
        \rho_{\infty}^b
        & = \frac{1}{4} \exp \left( -\frac{1}{b} - 2\sum_{j = 1}^{b-1}\frac{1}{j} \right) + \frac{1}{2}\int_0^1 \exp \left( \frac{y^b-1}{b} + 2\sum_{j = 1}^{b-1} \frac{y^j-1}{j} \right)dy \\
        & = \frac{1}{4} \exp \left(\frac{1}{b} - 2\sum_{j = 1}^b\frac{1}{j} \right) + \frac{1}{2}\int_0^1 \exp \left( -2\sum_{j = 1}^b \frac{1-y^j}{j} \right) \exp \left( \frac{1-y^b}{b} \right)dy.
    \end{aligned}
\end{equation}
Now the procedure is analogous to the one appearing in \cite{Gonzalez1974}. First note
\begin{align*}
    \sum_{j=1}^{b} \frac{1-y^j}{j}
    & = \sum_{j=1}^{b} \int_y^1 t^{j-1}\,dt = \int_y^1 \sum_{j=1}^{b} t^{j-1}\,dt = \int_y^1 \frac{1-t^b}{1-t}\,dt \\
    & = \begin{bmatrix}x=b(1-t)\\dx=-b\,dt\end{bmatrix} = \int_0^{b(1-y)} \frac{1-(1-\frac{x}{b})^b}{x}\,dx,
\end{align*}
and therefore
\begin{align*}
	2b\cdot\rho_{\infty}^b
	& = \frac{b}{2} \exp \left(\frac{1}{b} - 2\sum_{j = 1}^{b}\frac{1}{j} \right) + b\int_0^1 \exp \left( -2\sum_{j = 1}^b \frac{1-y^j}{j} \right) \exp \left( \frac{1-y^b}{b} \right) \, dy \\
	& = \frac{1}{2} \exp \left( \ln b + \frac{1}{b} - 2\sum_{j = 1}^{b}\frac{1}{j} \right) \\
	& \qquad + \int_0^1 \exp\left(-2 \int_0^{b(1-y)} \frac{1-(1-\frac{x}{b})^b}{x}\,dx \right) \exp\left(\frac{1-y^b}{b}\right)b\,dy\\
	& = \begin{bmatrix}\tilde{y}=b(1-y)\\d\tilde{y}=-b\,dy\end{bmatrix} = \frac{1}{2} \exp \left[ \ln b + \frac{1}{b} - 2\left(\gamma + \ln b + \frac{1}{2b} + O\left( \frac{1}{b^2} \right) \right) \right] \\
        & \qquad + \int_0^b \exp\left(-2 \int_0^{\tilde{y}} \frac{1-(1-\frac{x}{b})^b}{x}\,dx \right) \exp\left( \frac{1 - \left( 1 - \frac{\tilde{y}}{b} \right)^b}{b} \right) \,d\tilde{y},
\end{align*}
where we used the standard approximation of the harmonic number $\sum_{j = 1}^b \frac{1}{j}$ (and $\gamma =0.5772...$ is the Euler–Mascheroni constant). The dominated convergence theorem now implies
\begin{equation}
    \lim_{b\to\infty}	2b\cdot\rho_{\infty}^{b}
	= \int_0^\infty \exp\left[-2\int_0^{y} \frac{1-e^{-x}}{x} dx\right]dy
	= 0.7475979202\dots
\end{equation}
Notice that this is precisely the R\'{e}nyi's car parking constant from \eqref{eq:Renyis-cp-const}.

Next, we show $\lim_{b\to\infty} 2b\cdot\rho_{\star}^{b}=1$. From \eqref{eq:rho_via_y} it follows that for each $b\in\bbN$
\begin{equation}\label{eq:2brho^-1}
	\frac{1}{2b\cdot\rho_{\star}^{b}} = \frac{2y_\star^2}{b(1-y_\star^4)}+\frac{1-2y_\star^{2b}}{1-y_\star^{2b}},
\end{equation}
where $y_\star$ is the unique positive number $0<y_\star<1$ (depending on $b$) which solves \eqref{eq:ystar}. Note that
$$y_\star^{2b}=\frac{1}{1+2y_\star^2+2y_\star^4+\dots+2y_\star^{2b-2}+y_\star^{2b}}<\frac{1}{2b\cdot y_\star^{2b}},$$
hence,
$$\left(y_\star^{2b}\right)^2 < \frac{1}{2b},$$
and therefore
\begin{equation}\label{eq:limit_1}
	\lim_{b\to\infty} y_\star^{2b} = 0.
\end{equation}
Now, by letting $b\to\infty$ in \eqref{eq:ystar} we get
$$\lim_{b\to\infty} y_\star^{2b}\frac{1 - y_\star^{2b}}{1-y_\star^2}(1 + y_\star^2)=\frac{0}{\lim_{b\to\infty} 1-y_\star^2} = 1,$$
and therefore $\lim_{b\to\infty} 1-y_\star^2=0$, thus
\begin{equation}\label{eq:limit_2}
	\lim_{b\to\infty} y_\star=1.
\end{equation}
Lastly, from \eqref{eq:limit_1} we get
\begin{equation}\label{eq:limit_3}
	\lim_{b\to\infty} b(1-y_\star) = \lim_{b\to\infty} -b(y_\star-1) = \lim_{b\to\infty} -b\ln y_\star = \lim_{b\to\infty} \ln\frac1{\sqrt{y_\star^{2b}}} = +\infty.
\end{equation}

We can now substitute the limits obtained in \eqref{eq:limit_1}, \eqref{eq:limit_2}, and \eqref{eq:limit_3} into \eqref{eq:2brho^-1} and obtain
\begin{equation*}
	\lim_{b\to\infty} \frac{1}{2b\cdot\rho_{\star}^{b}} = \lim_{b\to\infty} \left( \frac{2y_\star^2}{b(1-y_\star)(1+y_\star)(1+y_\star^2)}+\frac{1-2y_\star^{2b}}{1-y_\star^{2b}}
	\right)=1,
\end{equation*}
thus proving
$$\lim_{b\to\infty} 2b\cdot\rho_{\star}^{b}=1.$$

\section{Conclusions}\label{sec:conclusion}
In this paper we analyze the model of Rydberg atoms on a two row square ladder.
This model is a direct generalization of the nearest neighbor exclusion model
(known in the literature as Flory's model, see
\cite{Fan1992, Baram1992}). Due to their large size, modeled by the
parameter of the model $b \ge 1$, referred to as the blockade range,
the atoms excited to a Rydberg state can cause much more than just the
nearest neighbor exclusion. We analyze both the dynamic and the equilibrium
variant of the model.

In the dynamic version, all the atoms are in the neutral state at the beginning of the process, and are then excited to a Rydberg state randomly and sequentially until a jammed configuration (in which no more neutral atoms can be excited to a Rydberg state without violating the constraint imposed by the blockade range) is reached. We derive a closed formula for the expected density of excited atoms, i.e.\ jamming limit, for an arbitrary blockade range $b \ge 1$. This result can be interpreted in terms of unfriendly seating arrangement at a dinning table (see \cite{Chern2015}), or unfriendly theater seating arrangement problem (see \cite{Georgiou2009}) with a higher level of unfriendliness.
The results can be also relevant for modeling non-pharmaceutical measures for
fighting epidemics, as witnessed during the recent COVID outbreak.

In the equilibrium variant, where the space of all jammed configurations is
explored with a flat measure, we compute the complexity function for arbitrary
$b \ge 1$. This, again, generalizes the results known for the Flory's model,
in which only the nearest neighbors are excluded (see \cite{Puljiz2023b}).
As a counterpart of the jamming limit, we compute the argument of the maximum
of the complexity functions for all $b\ge 1$, and we compare those arguments
with the jamming limits obtained for the dynamic model with the same value
of $b$. Beside noticing obvious violation of the Edwards hypothesis, we
analyze this difference further, and show that the behavior of the two models
is radically different as $b$ tends to infinity.

One possible way to generalize the obtained results is to consider graphs similar to the ladder shown in Figure \ref{fig:ladder}, but consisting of more than just two rows. These kind of graphs were considered in \cite{Georgiou2009}, and were referred to as the theater consisting of $m \ge 2$ rows of seats. The authors there study only the nearest neighbor exclusion process (i.e.\ they set $b = 1$), and they stress that they do not know how to calculate the jamming limit for $m > 2$ analytically, providing, instead, the Monte Carlo approximations for values of $m$ between $3$ and $15$. They also comment that one could add links from nodes in the bottom row to nodes in the top row to form a torus. If such a modification is performed, the case $m = 3$ becomes easier to study (as already observed in \cite{Baram1994}). The biggest problem with that kind of graphs for bigger values of $m$ is that, after an atom is excited to a Rydberg state, the graph does not necessarily decompose into two separate subgraphs. The situation here becomes simpler, though, when one considers Rydberg models with bigger blockade range $b$. If $b$ is big enough, each excited atom will separate (with its blockage effect) the original graph into two subgraphs that evolve independently. The efforts to extend our results in this direction are currently underway. Another natural thing to do would be to consider non-integral blockade ranges expressed in terms of the Euclidean distance. For the ladder graphs, such relaxations could lead to more tractable models. For example, $b = \sqrt{2}$ would yield, essentially, a one-dimensional model. It would also be interesting to consider these problems in finite portions of the regular hexagonal lattice.

\section*{Acknowledgments}
\noindent
We wish to thank Professor Pavel Krapivsky for introducing us to Rydberg models, and for bringing to our attention references \cite{Fan1992, Baram1992}. T.\ Do\v{s}li\'c gratefully acknowledges partial support by the Slovenian ARIS (program P1-0383, grant no.\ J1-3002) and by COST action CA21126 NanoSpace. Financial support of the Croatian Science Foundation (project IP-2022-10-2277) is gratefully acknowledged by S.\ Šebek.

\bibliographystyle{iopart-num}
\bibliography{../../settlement_planning_literature}

\providecommand{\newblock}{}
\begin{thebibliography}{10}
\expandafter\ifx\csname url\endcsname\relax
  \def\url#1{{\tt #1}}\fi
\expandafter\ifx\csname urlprefix\endcsname\relax\def\urlprefix{URL }\fi
\providecommand{\eprint}[2][]{\url{#2}}

\bibitem{Evans1993}
Evans J~W 1993 {\em Reviews of Modern Physics\/} {\bf 65} 1281--1329 ISSN
  1539-0756

\bibitem{Talbot2000}
Talbot J, Tarjus G, Van~Tassel P~R and Viot P 2000 {\em Colloids and Surfaces
  A: Physicochemical and Engineering Aspects\/} {\bf 165} 287--324 ISSN
  0927-7757

\bibitem{Torquato2002}
Torquato S 2002 {\em Random heterogeneous materials\/} [corr. 2. print.] ed
  ({\em Interdisciplinary applied mathematics\/} no~16) (New York [u.a.]:
  Springer) ISBN 0387951679 includes bibliographical references and index

\bibitem{Adamczyk2017}
Adamczyk Z (ed) 2017 {\em Particles at interfaces\/} second edition ed ({\em
  Interface science and technology\/} no volume 20) (London: Academic Press)
  ISBN 0081012489

\bibitem{Bressloff2013}
Bressloff P~C and Newby J~M 2013 {\em Reviews of Modern Physics\/} {\bf 85}
  135--196 ISSN 1539-0756

\bibitem{Chen2002}
Chen J~Y, Klemic J~F and Elimelech M 2002 {\em Nano Letters\/} {\bf 2} 393--396
  ISSN 1530-6992

\bibitem{Elimelech2003}
Elimelech M, Chen J~Y and Kuznar Z~A 2003 {\em Langmuir\/} {\bf 19} 6594--6597
  ISSN 1520-5827

\bibitem{Gray2006}
Gray J~L, Hull R and Floro J~A 2006 {\em Journal of Applied Physics\/} {\bf
  100} ISSN 1089-7550

\bibitem{Katsman2013}
Katsman A, Beregovsky M and Yaish Y~E 2013 {\em Journal of Applied Physics\/}
  {\bf 113} ISSN 1089-7550

\bibitem{Torquato2010}
Torquato S and Stillinger F~H 2010 {\em Reviews of Modern Physics\/} {\bf 82}
  2633--2672 ISSN 1539-0756

\bibitem{Flory1939}
Flory P~J 1939 {\em Journal of the American Chemical Society\/} {\bf 61}
  1518--1521 ISSN 1520-5126

\bibitem{Renyi1958}
R{\'{e}}nyi A 1958 {\em Publ. Math. Inst. Hung. Acad. Sci.\/} {\bf 3} 109--127
  ISSN 0541-9514

\bibitem{D'Orsogna2007}
D{\textquoteright}Orsogna M~R, Chou T and Antal T 2007 {\em Journal of Physics
  A: Mathematical and Theoretical\/} {\bf 40} 5575--5584 ISSN 1751-8121

\bibitem{Krapivsky2010}
Krapivsky P~L and Mallick K 2010 {\em Journal of Statistical Mechanics: Theory
  and Experiment\/} {\bf 2010} P07007 ISSN 1742-5468

\bibitem{Privman1992}
Privman V 1992 {\em Physical Review Letters\/} {\bf 69} 3686--3688 ISSN
  0031-9007

\bibitem{Krapivsky1994}
Krapivsky P~L 1994 {\em Journal of Statistical Physics\/} {\bf 74} 1211--1225
  ISSN 1572-9613

\bibitem{DeSmedt2002}
De~Smedt G, Godr{\`{e}}che C and Luck J~M 2002 {\em The European Physical
  Journal B - Condensed Matter\/} {\bf 27} 363--380 ISSN 1434-6036

\bibitem{Saffman2010}
Saffman M, Walker T~G and M{\o}lmer K 2010 {\em Reviews of Modern Physics\/}
  {\bf 82} 2313--2363 ISSN 1539-0756

\bibitem{Jaksch2000}
Jaksch D, Cirac J~I, Zoller P, Rolston S~L, C{\^{o}}t{\'{e}} R and Lukin M~D
  2000 {\em Physical Review Letters\/} {\bf 85} 2208--2211 ISSN 1079-7114

\bibitem{Liebisch2005}
Liebisch T~C, Reinhard A, Berman P~R and Raithel G 2005 {\em Physical Review
  Letters\/} {\bf 95} 253002 ISSN 1079-7114

\bibitem{Viteau2012}
Viteau M, Huillery P, Bason M~G, Malossi N, Ciampini D, Morsch O, Arimondo E,
  Comparat D and Pillet P 2012 {\em Physical Review Letters\/} {\bf 109} 053002
  ISSN 1079-7114

\bibitem{Malossi2014}
Malossi N, Valado M~M, Scotto S, Huillery P, Pillet P, Ciampini D, Arimondo E
  and Morsch O 2014 {\em Physical Review Letters\/} {\bf 113} 023006 ISSN
  1079-7114

\bibitem{Sanders2014}
Sanders J, van Bijnen R, Vredenbregt E and Kokkelmans S 2014 {\em Physical
  Review Letters\/} {\bf 112} 163001 ISSN 1079-7114

\bibitem{Bernien2017}
Bernien H, Schwartz S, Keesling A, Levine H, Omran A, Pichler H, Choi S, Zibrov
  A~S, Endres M, Greiner M, Vuleti{\'{c}} V and Lukin M~D 2017 {\em Nature\/}
  {\bf 551} 579--584 ISSN 1476-4687

\bibitem{Turner2018}
Turner C~J, Michailidis A~A, Abanin D~A, Serbyn M and Papi{\'{c}} Z 2018 {\em
  Nature Physics\/} {\bf 14} 745--749 ISSN 1745-2481

\bibitem{Ho2019}
Ho W~W, Choi S, Pichler H and Lukin M~D 2019 {\em Physical Review Letters\/}
  {\bf 122} 040603 ISSN 1079-7114

\bibitem{Khemani2019}
Khemani V, Laumann C~R and Chandran A 2019 {\em Physical Review B\/} {\bf 99}
  161101 ISSN 2469-9969

\bibitem{Marvel1938}
Marvel C~S and Levesque C~L 1938 {\em Journal of the American Chemical
  Society\/} {\bf 60} 280--284 ISSN 1520-5126

\bibitem{Nord1991}
Nord R~S 1991 {\em Journal of Statistical Computation and Simulation\/} {\bf
  39} 231--240 ISSN 1563-5163

\bibitem{Tory1983}
Tory E~M, Jodrey W~S and Pickard D~K 1983 {\em Journal of Theoretical
  Biology\/} {\bf 102} 439--445 ISSN 0022-5193

\bibitem{Wang2000}
Wang J~S 2000 {\em Colloids and Surfaces A: Physicochemical and Engineering
  Aspects\/} {\bf 165} 325--343 ISSN 0927-7757

\bibitem{Georgiou2009}
Georgiou K, Kranakis E and Krizanc D 2009 {\em Discrete Mathematics\/} {\bf
  309} 5120--5129 ISSN 0012-365X

\bibitem{Krapivsky2010a}
Krapivsky P~L, Redner S and Ben-Naim E 2010 {\em A Kinetic View of Statistical
  Physics\/} (Cambridge University Press) ISBN 9780511780516

\bibitem{Fan1992}
Fan Y and Percus J~K 1992 {\em Journal of Statistical Physics\/} {\bf 66}
  263--271 ISSN 1572-9613

\bibitem{Baram1992}
Baram A and Kutasov D 1992 {\em Journal of Physics A: Mathematical and
  General\/} {\bf 25} L493--L498 ISSN 1361-6447

\bibitem{Chern2015}
Chern H~H, Hwang H~K and Tsai T~H 2015 {\em Advances in Applied Mathematics\/}
  {\bf 65} 38--64 ISSN 0196-8858

\bibitem{Crisanti2000}
Crisanti A, Ritort F, Rocco A and Sellitto M 2000 {\em The Journal of Chemical
  Physics\/} {\bf 113} 10615--10634 ISSN 1089-7690

\bibitem{Dean2000}
Dean D~S 2000 {\em The European Physical Journal B\/} {\bf 15} 493--498 ISSN
  1434-6028

\bibitem{Lefevre2001}
Lef{\`{e}}vre A and Dean D~S 2001 {\em The European Physical Journal B\/} {\bf
  21} 121--128 ISSN 1434-6028

\bibitem{Lefevre2001a}
Lef{\`{e}}vre A and Dean D~S 2001 {\em Journal of Physics A: Mathematical and
  General\/} {\bf 34} L213--L220 ISSN 1361-6447

\bibitem{Palmer1985}
Palmer R~G and Frisch H~L 1985 {\em Journal of Statistical Physics\/} {\bf 38}
  867--872 ISSN 1572-9613

\bibitem{Elskens1987}
Elskens Y and Frisch H~L 1987 {\em Journal of Statistical Physics\/} {\bf 48}
  1243--1248 ISSN 1572-9613

\bibitem{Puljiz2023b}
Puljiz M, {{\v{S}}}ebek S and {{\v{Z}}}ubrini{\'{c}} J 2023 Complexity function
  for a variant of {F}lory model on a ladder {\em Proceedings of the 4th
  Croatian Combinatorial Days\/} CroCoDays (University of Zagreb Faculty of
  Civil Engineering) pp 93--109 ISBN 978-953-8168-63-5
  \urlprefix\url{https://www.grad.hr/crocodays/proc_ccd4/Puljiz.pdf}

\bibitem{Krapivsky2023}
Krapivsky P~L and Luck J~M 2023 {\em Journal of Physics A: Mathematical and
  Theoretical\/} {\bf 56} 255001 ISSN 1751-8121

\bibitem{Baule2018}
Baule A, Morone F, Herrmann H~J and Makse H~A 2018 {\em Reviews of Modern
  Physics\/} {\bf 90} 015006 ISSN 1539-0756

\bibitem{Friedman1964}
Friedman H~D, Shepp L, Rothman D and MacKenzie J~K 1964 {\em SIAM Review\/}
  {\bf 6} 180--182 ISSN 0036-1445

\bibitem{Krapivsky2020}
Krapivsky P~L 2020 {\em Physical Review E\/} {\bf 102} 062108 ISSN 2470-0053

\bibitem{Mackenzie1962}
Mackenzie J~K 1962 {\em The Journal of Chemical Physics\/} {\bf 37} 723--728
  ISSN 1089-7690

\bibitem{Doslic2024a}
Do{\v{s}}li{\'{c}} T, Puljiz M, {{\v{S}}}ebek S and {{\v{Z}}}ubrini{\'{c}} J
  2024 {\em Ars Mathematica Contemporanea\/} ISSN 1855-3966

\bibitem{Kranakis2010}
Kranakis E and Krizanc D 2010 {\em The Urinal Problem\/} (Springer Berlin
  Heidelberg) pp 284--295 ISBN 9783642131226

\bibitem{Luck2024}
Luck J~M 2024 {\em Journal of Physics A: Mathematical and Theoretical\/} {\bf
  57} 225002 ISSN 1751-8121

\bibitem{Page1959}
Page E~S 1959 {\em Journal of the Royal Statistical Society Series B:
  Statistical Methodology\/} {\bf 21} 364--374 ISSN 1467-9868

\bibitem{Gonzalez1974}
Gonz{\'{a}}lez J~J, Hemmer P~C and H{\o}ye J~S 1974 {\em Chemical Physics\/}
  {\bf 3} 228--238 ISSN 0301-0104

\bibitem{Doslic2019}
Do{\v{s}}li{\'{c}} T 2019 {\em Ars Mathematica Contemporanea\/} {\bf 17} 79--88
  ISSN 1855-3966

\bibitem{Baram1994}
Baram A and Kutasov D 1994 {\em Journal of Physics A: Mathematical and
  General\/} {\bf 27} 3683--3687 ISSN 1361-6447

\end{thebibliography}

\end{document}